\documentclass[12pt]{article}
\pdfoutput=1
\usepackage[colorlinks,linkcolor=Blue,citecolor=Blue,bookmarks,bookmarksnumbered]{hyperref}
\usepackage[scaled=0.85]{helvet}
\usepackage{amsmath,amssymb,accents,mathrsfs,XoohmE}
\usepackage{graphicx,color}
\usepackage{booktabs}
\usepackage{multirow}
\usepackage{placeins}
\usepackage{amsmath}

\definecolor{Green}  {rgb}{0.10,0.70,0.10} 
\definecolor{Orange} {rgb}{1.00,0.50,0.15} 
\definecolor{Red}    {rgb}{0.90,0.00,0.12} 
\definecolor{Purple} {rgb}{0.50,0.25,0.55} 
\definecolor{Turque} {rgb}{0.00,0.65,0.85} 
\definecolor{Blue}   {rgb}{0.00,0.00,1.00} 
\definecolor{Magenta}{rgb}{1.00,0.00,1.00} 
\definecolor{Gold}   {rgb}{1.00,0.75,0.25} 
\definecolor{Seaweed}{rgb}{0.01,0.24,0.09} 
\definecolor{Brown}  {rgb}{0.43,0.26,0.32} 
\definecolor{grey1}  {rgb}{0.20,0.20,0.20} 
\definecolor{grey2}  {rgb}{0.40,0.40,0.40} 
\definecolor{grey3}  {rgb}{0.60,0.60,0.60} 
\definecolor{grey4}  {rgb}{0.80,0.80,0.80} 
\definecolor{grey5}  {rgb}{0.90,0.90,0.90} 
\def\C#1#2{{\ifcase#1\or
             \color{Green}\or \color{Orange}\or \color{Red}\or
              \color{Purple}\or \color{Turque}\or \color{Blue}\or
               \color{Magenta}\or \color{Gold}\or \color{Seaweed}\or
                \color{Brown}\or\color{grey1}\or\color{grey2}\or
                 \color{grey3}\else\color{grey4}\fi#2}}

\definecolor{Slate} {rgb}{0.00,0.45,0.55}

\definecolor{AdinkraGreen}{rgb}{0.10196079, 0.61176473, 0.21960784 }
\definecolor{AdinkraViolet}{rgb}{0.42352942, 0.15294118, 0.4509804 }
\definecolor{AdinkraOrange}{rgb}{0.89803922, 0.57647061, 0.27450982}
\definecolor{AdinkraRed}{rgb}{0.78431374, 0, 0.12156863}

\def\rD{\mbox{\textcolor{AdinkraRed}{${\rm D}_4$}}}

\def\rD{{\rm D}}
\def\rI{{\rm I}}
\def\rJ{{\rm J}}

\def\hi{{\hat\imath}}

\def\hk{{\hat{k}}}

\def\fracm#1#2{\hbox{\large{${\frac{{#1}}{{#2}}}$}}}

\def\vCent#1{\vcenter{\hbox{\hss#1\hss}}}
\def\be{\begin{equation}}
\def\ee{\end{equation}}
\newcommand{\bea}{\begin{eqnarray}}
\newcommand{\eea}{\end{eqnarray}}
\newcommand{\ena}{\end{eqnarray}}

\def\pp{{\mathchoice
          {
              \kern 1pt%
              \raise 1pt
              \vbox{\hrule width5pt height0.4pt depth0pt
                    \kern -2pt
                    \hbox{\kern 2.3pt
                          \vrule width0.4pt height6pt depth0pt
                          }
                    \kern -2pt
                    \hrule width5pt height0.4pt depth0pt}%
                    \kern 1pt
           }
            {
              \kern 1pt%
              \raise 1pt
              \vbox{\hrule width4.3pt height0.4pt depth0pt
                    \kern -1.8pt
                    \hbox{\kern 1.95pt
                          \vrule width0.4pt height5.4pt depth0pt
                          }
                    \kern -1.8pt
                    \hrule width4.3pt height0.4pt depth0pt}%
                    \kern 1pt
            }
            {
              \kern 0.5pt%
              \raise 1pt
              \vbox{\hrule width4.0pt height0.3pt depth0pt
                    \kern -1.9pt  
                    \hbox{\kern 1.85pt
                          \vrule width0.3pt height5.7pt depth0pt
                          }
                    \kern -1.9pt
                    \hrule width4.0pt height0.3pt depth0pt}%
                    \kern 0.5pt
            }
            {
              \kern 0.5pt%
              \raise 1pt
              \vbox{\hrule width3.6pt height0.3pt depth0pt
                    \kern -1.5pt
                    \hbox{\kern 1.65pt
                          \vrule width0.3pt height4.5pt depth0pt
                          }
                    \kern -1.5pt
                    \hrule width3.6pt height0.3pt depth0pt}%
                    \kern 0.5pt
            }
        }}

\def\mm{{\mathchoice
                       {
                             \kern 1pt
               \raise 1pt    \vbox{\hrule width5pt height0.4pt depth0pt
                                  \kern 2pt
                                  \hrule width5pt height0.4pt depth0pt}
                             \kern 1pt}
                       {
                            \kern 1pt
               \raise 1pt \vbox{\hrule width4.3pt height0.4pt depth0pt
                                  \kern 1.8pt
                                  \hrule width4.3pt height0.4pt depth0pt}
                             \kern 1pt}
                       {
                            \kern 0.5pt
               \raise 1pt
                            \vbox{\hrule width4.0pt height0.3pt depth0pt
                                  \kern 1.9pt
                                  \hrule width4.0pt height0.3pt depth0pt}
                            \kern 1pt}
                       {
                           \kern 0.5pt
             \raise 1pt  \vbox{\hrule width3.6pt height0.3pt depth0pt
                                  \kern 1.5pt
                                  \hrule width3.6pt height0.3pt depth0pt}
                           \kern 0.5pt}
                       }}

\def\ad{{\kern0.5pt
                   \alpha \kern-5.05pt \raise5.8pt\hbox{$\textstyle.$}\kern
0.5pt}}

\def\bd{{\kern0.5pt
                   \beta \kern-5.05pt \raise5.8pt\hbox{$\textstyle.$}\kern
0.5pt}}

\def\qd{{\kern0.5pt
                   q \kern-5.05pt \raise5.8pt\hbox{$\textstyle.$}\kern
0.5pt}}
\def\Dot#1{{\kern0.5pt
     {#1} \kern-5.05pt \raise5.8pt\hbox{$\textstyle.$}\kern
0.5pt}}

\catcode`@=11
\def\un#1{\relax\ifmmode\@@underline#1\else
        $\@@underline{\hbox{#1}}$\relax\fi}
\catcode`@=12

\def\a{\alpha}
\def\b{\beta}
\def\c{\chi}
\def\d{\delta}
\def\e{\epsilon}

\def\g{\gamma}

\def\i{\iota}
\def\j{\psi}
\def\k{\kappa}
\def\l{\lambda}
\def\m{\mu}

\def\o{\omega}

\def\r{\rho}
\def\s{\sigma}
\def\t{\tau}

\def\G{\Gamma}

\def\L{\Lambda}
\def\O{\Omega}
\def\P{\Pi}

\def\S{\Sigma}

\def\ca{{\cal A}}

\def\cf{{\cal F}}

\def\dslash{\not{\hbox{\kern-2pt $\partial$}}}
\def\Dslash{\not{\hbox{\kern-4pt $D$}}}
\def\pslash{\not{\hbox{\kern-2.3pt $p$}}}
 \newtoks\slashfraction
 \slashfraction={.13}
 \def\slash#1{\setbox0\hbox{$ #1 $}
 \setbox0\hbox to \the\slashfraction\wd0{\hss \box0}/\box0 }
 
\def\kcr{{\hbox{\ro \char'170}}}                
\def\ktl{{\hbox{\ro \char'170}}}        
\def\ktr{{\hbox{\ro \char'170}}}        
\def\kbl{{\hbox{\ro \char'170}}}        
\def\kbr{{\hbox{\ro \char'170}}}        

\def\plpl{\raise-2pt\hbox{$\raise3pt\hbox{$_+$}\hskip-6.67pt\raise0.0pt
\hbox{$^+$}\hskip 0.01pt$}}
\def\mimi{\raise-2pt\hbox{$\raise3pt\hbox{$_-$}\hskip-6.67pt\raise0.0pt
\hbox{$^-$}\hskip 0.01pt$}} 

\def\bo{{\raise.15ex\hbox{\large$\Box$}}}               
\def\pa{\partial}                                       
\def\TH{{\raise.2ex\hbox{$\displaystyle \bigodot$}\mskip-4.7mu \llap H \;}}
\def\face{{\raise.2ex\hbox{$\displaystyle \bigodot$}\mskip-2.2mu \llap {$\ddot
        \smile$}}}                                      

\def\dt#1{\on{\hbox{\bf .}}{#1}}                
\def\Dot#1{\dt{#1}}

\def\Tilde#1{\widetilde{#1}}                    
\def\Hat#1{\widehat{#1}}                        
\def\leftrightarrowfill{$\mathsurround=0pt \mathord\leftarrow \mkern-6mu
        \cleaders\hbox{$\mkern-2mu \mathord- \mkern-2mu$}\hfill
        \mkern-6mu \mathord\rightarrow$}
\def\dvec#1{\vbox{\ialign{##\crcr
        \leftrightarrowfill\crcr\noalign{\kern-1pt\nointerlineskip}
        $\hfil\displaystyle{#1}\hfil$\crcr}}}           
\def\dt#1{{\buildrel {\hbox{\LARGE .}} \over {#1}}}     

\def\fracm#1#2{\hbox{\large{${\frac{{#1}}{{#2}}}$}}}
\def\sfrac#1#2{{\vphantom1\smash{\lower.5ex\hbox{\small$#1$}}\over
        \vphantom1\smash{\raise.4ex\hbox{\small$#2$}}}} 
\def\bfrac#1#2{{\vphantom1\smash{\lower.5ex\hbox{$#1$}}\over
        \vphantom1\smash{\raise.3ex\hbox{$#2$}}}}       
\def\afrac#1#2{{\vphantom1\smash{\lower.5ex\hbox{$#1$}}\over#2}}    

\def\pa{\partial}      
\let\bm\relax
\newcommand{\bm}[1]{{\boldsymbol{#1}}}

\def\ad{{\dot{\alpha}}}
\def\bd{{\dot{\beta}}}

 \font\rOpe=cmsy10                        
 \def\ktl{{\hbox{\rOpe\char'170}}}        
 \def\kbl{{\hbox{\rOpe\char'170}}}        
 \def\kcr{{\reflectbox{\rOpe\char'170}}}        
 \def\ktr{{\reflectbox{\rOpe\char'170}}}        
 \def\kbr{{\reflectbox{\rOpe\char'170}}}        
 \def\Border{\vbox{\hsize0pt
        \setlength{\unitlength}{1mm}
        \newcount\xco
        \newcount\yco
        \xco=-21
        \yco=12
        \begin{picture}(0,0)(-7.5,0)
        \put(\xco,\yco){$\ktl$}
        \advance\yco by-1
        {\loop
        \put(\xco,\yco){$\kcr$}
        \advance\yco by-2
        \ifnum\yco>-240
        \repeat
        \put(\xco,\yco){$\kbl$}}
        \xco=170
        \yco=12
        \put(\xco,\yco){$\ktr$}
        \advance\yco by-1
        {\loop
        \put(\xco,\yco){$\kcr$}
        \advance\yco by-2
        \ifnum\yco>-240
        \repeat
        \put(\xco,\yco){$\kbr$}}
        \put(-19.5,13){\scalebox{.6065}{%
         University of Maryland Center for String and Particle  Theory \&\ Physics Department%
        |University of Maryland Center for String and Particle  Theory \&\ Physics Department}}
        \put(-19.5,-241.5){\scalebox{.5835}{%
         ****University of Maryland * Center for String and
         Particle  Theory* Physics Department****University of Maryland *Center
        for String and Particle  Theory* Physics Department}}
        \end{picture}
        \par\vskip-8mm}}
\definecolor{UMred}{rgb}{.9,.05,.2}
\definecolor{HUblue}{rgb}{.0,.3,.7}
 \def\UMbanner{\vbox{\hsize0pt
        \setlength{\unitlength}{.4mm}
        \thicklines  
        \begin{picture}(0,0)(-30,-10)
        \put(165,2){\line(1,0){4}}
        \put(170,2){\line(1,0){4}}
        \put(180,2){\line(1,0){4}}
        \put(175,-14){\line(1,0){4}}
        \put(180,-14){\line(1,0){4}}
        \put(185,-14){\line(1,0){4}}
        \put(169,-14){\line(0,1){16}}
        \put(170,-14){\line(0,1){16}}
        \put(179,-14){\line(0,1){16}}
        \put(180,-14){\line(0,1){16}}
        \put(184,-14){\line(0,1){16}}
        \put(185,-14){\line(0,1){16}}
        \put(169,2){\oval(8,32)[bl]}
        \put(170,2){\oval(8,32)[br]}
        \put(179,-14){\oval(8,32)[tl]}
        \put(185,-14){\oval(8,32)[tr]}
        \end{picture}
        \par\vskip-6.5mm
        \thicklines}}

\definecolor{Red}    {rgb}{0.90,0.00,0.12} 
\definecolor{Blue}   {rgb}{0.00,0.00,1.00} 
\definecolor{Green}  {rgb}{0.10,0.70,0.10} 
\definecolor{Turque} {rgb}{0.00,0.65,0.85} 
\definecolor{Orange} {rgb}{1.00,0.50,0.15} 
\definecolor{Magenta}{rgb}{1.00,0.00,1.00} 
\definecolor{Gold}   {rgb}{1.00,0.75,0.25} 
\definecolor{Seaweed}{rgb}{0.01,0.24,0.09} 
\definecolor{Purple} {rgb}{0.50,0.25,0.55} 
\definecolor{Brown}  {rgb}{0.43,0.26,0.32} 
\definecolor{grey1}  {rgb}{0.20,0.20,0.20} 
\definecolor{grey2}  {rgb}{0.40,0.40,0.40} 
\definecolor{grey3}  {rgb}{0.60,0.60,0.60} 
\definecolor{grey4}  {rgb}{0.80,0.80,0.80} 
\definecolor{grey5}  {rgb}{0.90,0.90,0.90} 
\def\C#1#2{{\ifcase#1\or
             \color{Red}\or \color{Green}\or \color{Blue}\or\
              \color{Turque}\or \color{Orange}\or \color{Magenta}\or 
               \color{Gold}\or \color{Seaweed}\or \color{Purple}\or
                \color{Brown}\or\color{grey1}\or\color{grey2}\or
                 \color{grey3}\else\color{grey4}\fi#2}}

\definecolor{Slate} {rgb}{0.00,0.45,0.55}

\newdimen\parshift\parshift=\parindent
\catcode`@=11
 \long\def\@footnotetext#1{\insert\footins{\reset@font\footnotesize
           \interlinepenalty\interfootnotelinepenalty\splittopskip%
            \footnotesep\splitmaxdepth\dp\strutbox\floatingpenalty\@MM%
             \hsize\columnwidth\addtolength{\hsize}{-2\parindent}
              \@parboxrestore\protected@edef\@currentlabel%
              {\csname p@footnote\endcsname\@thefnmark}%
                \color@begingroup%
                 \@makefntext{\rule\z@\footnotesep\ignorespaces#1%
                  \@finalstrut\strutbox}%
                \color@endgroup}}
 \long\def\@makefntext#1{\hglue\parshift%
           \vbox{\noindent\baselineskip=11pt plus.5pt minus.5pt\hb@xt@0em{\hss\@makefnmark\kern1pt}#1}}
\catcode`@=12

\newskip\humongous \humongous=0pt plus 1000pt minus 1000pt
\def\caja{\mathsurround=0pt}
\def\eqalign#1{\,\vcenter{\openup2\jot \caja
        \ialign{\strut \hfil$\displaystyle{##}$&$
        \displaystyle{{}##}$\hfil\crcr#1\crcr}}\,}
\newif\ifdtup

\makeatletter
\def\section{\@startsection{section}{1}{\z@}
        {3ex plus-1ex minus-.2ex}{1pt plus1pt}{\large\sf\bfseries\boldmath}}
\def\subsection{\@startsection{subsection}{2}{\z@}
         {1.5ex plus-1ex minus-.2ex}{0.01pt plus1pt}{\sf\slshape}}
\def\subsubsection{\@startsection{subsubsection}{3}{\z@}
          {1.5ex plus-1ex minus-.2ex}{0.01pt plus0.2pt}{\sf\boldmath}}
\def\paragraph{\@startsection{paragraph}{4}{\z@}
           {.75ex \@plus.5ex \@minus.2ex}{-2mm}{\sf\bfseries\boldmath}}
\makeatother

 \allowdisplaybreaks
 \seceq

\def\DDt#1{\accentset{\hbox{\LARGE.\kern-2pt.}}{#1}}	
\def\dt#1{\accentset{\hbox{\large.}}{#1}}	
\def\ddt#1{\accentset{\hbox{\large\kern.5pt.\kern-1pt.}}{#1}}	
\def\CLS{\text{\slshape CLS}}

\def\eX{\rlap{\raisebox{.35ex}{\kern.45ex\scriptsize\it=}}{\boldsymbol X}}
\def\eY{\rlap{\raisebox{.35ex}{\kern.12ex\scriptsize\it=}}{\boldsymbol Y}}
\def\EX{\rlap{\raisebox{.45ex}{\kern.425ex\scriptsize\it=}}{X}}
\def\EY{\rlap{\raisebox{.45ex}{\kern.1ex\scriptsize\it=}}{Y}}

\def\ad{{\dt\a}}
\def\bd{{\dt\b}}

\let\bs\boldsymbol
\def\rD{{\rm D}}

\def\bDb{\hbox{\kern2pt\vrule height10pt depth-9.2pt width6pt\kern-9pt{$\boldsymbol D$}}\mkern-2mu}

\def\bQb{\hbox{\kern2pt\vrule height10pt depth-9.2pt width6pt\kern-9pt{$\boldsymbol Q$}}}

\def\bK{{\bs{K}}}
\def\bL{{\bs{L}}}
\def\bM{{\bs{M}}}
\def\bN{{\bs{N}}}
\def\bU{{{\bs{U}}}}

\def\bV{{{\bs{V}}}}

\def\Br{{{\bs\rho}}}
\def\Bz{{{\bs\zeta}}}
\def\BB{{{\bs\beta}}}
\def\Br{{{\bs\rho}}}
\def\Bz{{{\bs\zeta}}}
\def\BB{{{\bs\beta}}}

\def\BSb{\hbox{\kern2.5pt\vrule height10pt depth-9.2pt width7pt\kern-10.25pt{$\boldsymbol{\mit\Sigma}$}}}

\def\rQb{\hbox{\kern1pt\vrule height10pt depth-9.2pt width6pt\kern-8pt{\bf Q}}}

\def\rBx#1#2{\hbox to#1{#2\hss}}

\begin{document}

\thispagestyle{empty}
\vbox{\Border\UMbanner}
\noindent{\small
\hfill{PP--017-020 \\ 
$~~~~~~~~~~~~~~~~~~~~~~~~~~~~~~~~~~~~~~~~~~~~~~~~~~~~~~~~~~~~$
$~~~~~~~~~~~~~~~~~~~~\,~~~~~~~~~~~~~~~~~~~~~~~~~\,~~~~~~~~~~~~~~~~$
 {HET-1711}
}
\vspace*{1mm}
\begin{center}
{\large \bf
On the Four Dimensional Holoraumy of the 
\\[6pt]  
4D, $\cal N$ = 1 Complex Linear Supermultiplet}   \\   [6mm]
{\large {
Wesley Caldwell\footnote{wcaldwell.email@gmail.com}$^{a}$, 
Alejandro N.\ Diaz\footnote{adiaz126@terpmail.umd.edu}$^{a}$, 
Isaac Friend\footnote{icfriend@uchicago.edu}$^{b}$,
S.\ James Gates, Jr.,\footnote{gatess@wam.umd.edu}$^{a, \, c}$,
Siddhartha Harmalkar\footnote{sharmalk@umd.edu}$^{a}$, 
Tamar Lambert-Brown\footnote{tlambert720@gmail.com}$^{a}$,
Daniel Lay\footnote{dlay@terpmail.com}$^{a}$, 
Karina Martirosova\footnote{kmartirosova95@gmail.com}$^{a}$, 
Victor A. Meszaros\footnote{victorameszaros@gmail.com}$^{a}$,
Mayowa Omokanwaye\footnote{mayowao@mit.edu}$^{d}$,
Shaina Rudman\footnote{shaina.rudman@gmail.com}$^{a}$, 
Daeljuck Shin\footnote{shindae.kr@gmail.com}$^{a}$, 
and Anthony Vershov\footnote{anthonyvershov@gmail.com}$^{a}$
}}
\\[6mm]
\emph{
\centering
$^a$Center for String and Particle Theory-Dept.\ of Physics,
University of Maryland, \\[-2pt]
4150 Campus Dr., College Park, MD 20472,  USA
\\[12pt] 
$^b$  University of Chicago, Department of Mathematics,
\\[1pt]
5734 S. University Ave., Chicago, IL, 60637 USA
\\[12pt]         
$^{c}$Department of Physics, Brown University,
\\[1pt]
Box 1843, 182 Hope Street, Barus \& Holley 545,
Providence, RI 02912, USA
\\[8pt] 
and
\\[8pt] 
$^d$ Massachusetts Institute of Technology, Department of Mathematics,
\\[1pt]
Cambridge, MA 02139-4307, USA
}
 \\*[5mm]
{ ABSTRACT}\\[4mm]
\parbox{142mm}{\parindent=2pc\indent\baselineskip=14pt plus1pt
We present arguments to support the existence of weight spaces for supersymmetric field 
theories and identify the calculations of information about supermultiplets to define such 
spaces via the concept of ``holoraumy.''  For the first time this is extended to the
complex linear superfield by a calculation of the commutator of supercovariant 
derivatives on all of its component fields.
 }
 \end{center}
\vfill
\vfill
\clearpage

\section{Introduction}

Sometimes simple questions have interesting answers.  One such simple 
question about supersymmetry in one dimension can be stated as follows:

$~~~~$  $~~~~$  $~~~~$ $~~$ $~~~~$ {\it {If one boson and one fermion 
are sufficient to define a system with}} 
\newline $~~~~$  $~~~~$  $~~~~$ $~~$  $~~~~$ 
{\it {one linearly realized supercharge, and if two bosons and two fermions}} 
\newline $~~~~$  $~~~~$  $~~~~$ $~~$  $~~~~$ 
{\it {are sufficient to define a system with two linearly realized supercharges,}} 
\newline $~~~~$  $~~~~$  $~~~~$ $~~$  $~~~~$ 
{\it {why cannot three bosons and three fermions be used to define a system}} 
\newline $~~~~$  $~~~~$  $~~~~$ $~~$  $~~~~$ 
{\it {with three linearly realized supercharges?}}

\noindent
Now one can, by trial-\&-error, attempt to succeed in carrying out the step 
described at the end of this question. But such attempts are doomed to fail.  What 
mathematical principle is responsible for forbidding this?

Among the most interesting supersymmetrical theories are those in ten dimensions:
\vskip1.0pt   $~~~~$  (a.) Yang-Mills theory,
\vskip1.0pt   $~~~~$  (b.) simple supergravity,
\vskip1.0pt   $~~~~$  (c.) type-IIA and type-IIB supergravities, \vskip1.0pt \noindent
and the theory of eleven dimensional supergravity.  All of these occur 
as the limits of the more complicated superstrings, heterotic strings, M-Theory, 
and F-Theory.  A common feature of these interesting supersymmetrical theories,
is that their spectrum is only known modulo a mass-shell condition, i.e. a dynamical
assumption.  This raises the question below.

$~~~~~~~~$  $~~~~~~$  $~~~~~$ $~~$ $~~~~$ {\it {Is it possible to discover the field
content spectrum 
of these}}  \newline
$~~~~~~~~$  $~~~~~~$  $~~~~~$ $~~$  $~~~~$ {\it {ten
and eleven dimensional supersymmetrical theories without}}
\newline
$~~~~~~~~$  $~~~~~~$  $~~~~~$ $~~$  $~~~~$ {\it {the use 
of dynamical assumptions?}}

We believe our two italicized questions are two manifestations of a gap in the 
current understanding of supersymmetry and highlight the challenge to establish 
a mathematical structure that possesses the ability to respond to these two questions 
on the basis of a principle.  While it is true all of these higher dimensional theories 
are complicated, far more than the one dimensional ones described in the first
paragraph, it has been our long and deeply-held suspicion that solving the first 
question should ultimately lead to an answer for the second.  Furthermore, we 
believe the source of these resolutions lies in finding the most primitive and 
elementary mathematical models of spacetime supersymmetry.

It was asserted in the work of \cite{KIAS}  that adinkras (as suggested previously 
\cite{adnk1}, and refined in a series of works \cite{adnkM1,adnkM2,codes1,codes2,codes3}
that firmly established them in the realms of the mathematics of decorated cubical
cohomology, graphs, coding theory, etc. \cite{GrphThry,YZ,WIAA}) are the most 
primitive mathematical models of spacetime supersymmetry.  On the basis of our research,
the answer to the first italicized question is the impossibility to construct three
real 3 $\times$ 3 matrices that provide \cite{GRana1,GRana2} ``L-matrices'' and 
``R-matrices''  ${\bm {\rm L}}{}^{(\cal R)}_\rI\,$ and ${\bm {\rm R}}{}^{(\cal R)}_\rI$ to 
satisfy the ``Garden Algebra'' conditions
\be { \eqalign{
{\bm {\rm L}}{}^{(\cal R)}_\rI  \, {\bm {\rm R}}{}^{(\cal R)}_\rJ ~+~ {\bm {\rm L
 }}{}^{(\cal R)}_\rJ \, {\bm {\rm R}}{}^{(\cal R)}_\rI  ~&=~ 2\,\d_{\rI\rJ}\,
{\bm {\rm I}}{}_{d \times d} ~~,\cr
{\bm {\rm  R}}{}^{(\cal R)}_\rI   \,  {\bm {\rm L}}{}^{(\cal R)}_\rJ ~+~
{\bm {\rm R}}{}^{(\cal R)}_\rJ   \,  {\bm {\rm L}}{}^{(\cal R)}_\rI    ~&=~ 2\,\d_{\rI\rJ}
 {\bm {\rm I}}{}_{d \times d} ~~,  \cr
~~~ {\bm {\rm  R}}{}^{(\cal R)}_\rI  ~=~ [ \,  {\bm {\rm L}}{}^{(\cal R)}_\rI \, ]{}^{-1}
&~~,
}}\label{GarDNAlg2}
\ee
where in this special case d = 3.  Additionally one should keep in mind that
L-matrices are essentially adjacency matrices for adinkra graphs 
in the representation ($\cal R$).

Our ultimate goal is to exploit the structure and eventual complete understanding 
of adinkras to attack and make definitive statements about the long unsolved 
problems about the spectrum of the interesting higher dimensional 
theories.

In a recent work \cite{adnkBillions}, there was presented evidence (for 
minimal supermultiplets that possess 4D, $\cal N$ = 1 supersymmetry), via 
use of results strictly in the realm of four dimensional calculations, that it is 
possible to define a putative 4D, $\cal N$ = 1 weight space.

In chapter two, we discuss the rationale for the approach to using adinkras
as the pathway to establish for supersymmetrical field theories a robust
representation theory that is analogous to that which exist for compact
Lie algebras.

In chapter three, we will give evidence, within the context of minimal 4D, 
$\cal N$ = 1 supermultiplets, that the concept of fermionic holoraumy
\cite{adnkholor1,adnkholor2} can play an important role of identifying 
``eigenvalues''  to organize a representation theory of these supermultiplets.  
Although it is mostly over looked, there are actually {\em {ten}} such 
representations and we discuss a weight space interpretation for all of 
them.

The fourth chapter is devoted to a review of the complex linear supermultiplet
\cite{CLShist1,CLShist2,CLShist3}.  The usual component field structure is presented 
together with the expressions for the superspace covariant derivative acting 
on each of the component fields that lie as the lowest component of a superfield.  

The new results we report occur in the fifth chapter with the calculation of the 
{\em {commutator}} of two superspace covariant derivative acting on each of 
the component fields that lie as the lowest component of a superfield.  Following 
previous presentations, this allows the first derivation of the holoraumy tensors 
associated with the complex linear supermultiplet.  It is shown there appear
new extensions of the 4D, $\cal N$ = 1 holoraumy tensors that were not
present in the case of the minimal 4D, $\cal N$ = 1 supermultiplet representations.

The sixth and seventh chapter consist of the 0 dimensional spatial (0-brane) reduction
results for the complex linear supermultiplet together with the process of field
redefinitions required to obtain a 0-brane formulation that is closely related
to a valise formulation of the adinkra shadow of the complex linear supermultiplet.

The eighth chapter is devoted to a descriptions of the codes that were used
as one of the alternate derivation paths to the results shown in chapter five.
These were also used as checks on sets of calculations that were undertaken
by hand with pencil and paper.  There is also provided a hyperlink that may
be used to download the codes described in this chapter.

In the final chapter, we present our conclusions.

\newpage
\section{What Is The Problem Adinkras Are Proposing To Solve?}
\label{s00a}

Lie algebras have had a tremendous impact in theoretical physics.  One point
of reference marking the increase in their physics relevance came about early 
in the 1960's when Ne'eman \cite{qYN},  Gell-Mann \cite{qGM}, and George Zweig 
\cite{qGZ1,qGZ2} all independently proposed the use of the group SU(3) to 
provide a mathematical foundation for the classification of nuclear matter.  
Later history would show these researchers discovered the most elementary 
physical as well as mathematical structures -``the quarks'' - upon which the 
current theory of quantum chromodynamics is based.  In this case, the search 
for ``elementarity'' as a key concept of the realization of the SU(3) symmetry 
was an important hint for the task of building the {\em {Standard}} {\em {Model}}.  
The Standard Model is the most comprehensive theory of Nature that has ever 
been constructed.  Currently, it is consistent with a minimum of tens of thousands 
of observations and in some of its domains provides predictions that match 
observations of Nature to better than one part in billion.

Independently of physics, the subject of Lie algebras in the mathematical literature, 
due to many mathematicians such as Sophus Lie, Felix Klein, Wilhelm Killing, Friedrich 
Engel, and most notably Elie Cartan, had evolved to a high level of understanding.
In his 1894 thesis Cartan \cite{CarT} was able to provide a rigorous classification.  Key concepts 
in this achievement are the notions of ``weight'' and ``root'' spaces which arise from 
the fact that all Lie algebras possess a Jordan-Chevalley \cite{Chev} decomposition.  For any Lie 
algebra, there exists a maximal set of mutually commuting matrices among 
those that represent all the generators.  These mutually commuting matrices possess 
eigenvalues with which one can classify all the representations of the group related 
to the algebra's generators.

For example, the number of states for a representation of SU(2) is specified by $j$ 
(where $j$ can be one-half times an odd integer, or any non-negative integer) with 
the number of states being given by 
\be  {
{\bm {N}}{}_{SU(2)}(j) ~=~ \left(  \,  2 \, j ~+~ 1   \, \right)  ~~~.
} \label{WyL1}
\ee
For physicists, the quantity $j$ arises as the absolute value of the eigenvalue of ``the
third component of the angular moment.''  In a similar manner, the number of states for 
a representation of SU(3) is specified by $p$ and $q$ (where $p$ and $q$ can be any 
non-negative integers) with the corresponding number of states being given by 
\be  {
{\bm {d}}{}_{SU(3)} (p, \, q)~=~\fracm 12 \,  \left(  \,  p ~+~ 1   \, \right)\, 
 \left(  \,  q ~+~ 1   \, \right)\,  \left(  \,  p ~+~ q ~+~ 2   \, \right)\, 
  ~~~.
} \label{WyL2}
\ee
Once more for physicists, the quantities $p$ and $q$ arise as the number of 
elementary quark triplets and anti-quark triplets that are used to make a composite
state.

The formulae (\ref{WyL1}) and (\ref{WyL2}) are related to special cases of the 
general formula
\be
{\rm {dim}}\, \G(w^h) ~=~ 2 \, \prod_{\mu \, > \, 0 } \,{ {\left[  \, \mu \, \cdot \,( \, w^h ~+~
\fracm 12 \mu^+ \,)   \, \right] }  \over  {\left[   \, \mu \, \cdot \,  \mu^+
 \   \, \right] } }
\ee
which was discovered by Hermann Weyl  \cite{WyL1,WyL2,WyL3}.  In this formula, 
$ \G(w^h)$ denotes a representation labeled by the highest weight vector $w^h$, 
the positive roots of the algebra are denoted by $\mu$, and the sum of all the positive 
roots is denoted by $\mu^+$.  This formula is based on the fact that the Jordan-Chevalley
decomposition is at the heart of the representation theory of Lie algebras.

Little in the context of spacetime supersymmetry compares to the comprehensive 
nature of the representation theory achieved for Lie algebras.  Two Casimir operators, 
the ``superspin'' and ``mass,'' are often used to provide a basis for classification.  But 
these provide at best a partial classification.  We have been developing the theory of 
adinkras to fill this gap.  In particular, we have been diligently exploring adinkra's 
mathematical structure in search of quantities that will play the role of $j$ for SU(2) 
or $p$ and $q$ in the case of SU(3) discussed above.

One might ask, ``Why is the representation theory of spacetime supersymmetry so 
distinctive when compared to that of usual Lie algebras?''  Our investigations 
indicate the failure of the Jordan-Chevalley decomposition is the cause of the 
disjuncture.  To illustrate this point, it suffices to carry out a side-by-side comparison 
between the su(3) Lie algebra\footnote{We are using the convention where the notation 
``su(3)'' refers to the Lie algebra while ``SU(3)'' refers to the Lie group.} and the case of 
one dimensional, $N$ = 2 supersymmetry.

In the case of the su(3) Lie algebra, we have eight hermitian 3 $\times$ 3
matrices which denote by ${\bm T}_{\Hat \a}$ ($=$ ${\bm T}_1$, ${\bm T
}_1$, $\dots$, ${\bm T}_8$) that satisfy the familiar commutator conditions
\be
\left[  \,  {\bm T}_{\Hat \a} ~,~  {\bm T}_{\Hat \b}  \, \right]   ~=~ i \, f {}_{\Hat \a}
{}_{\Hat \b}{}^{\Hat \g} \, {\bm T}_{\Hat \g}      ~~~,
\ee
where $f {}_{\Hat \a}{}_{\Hat \b}{}^{\Hat \g}$ are the well-known structure constants
for the su(3) Lie algebra.  Using these structure constants, one can show that ${\bm 
 T}_3$ and  ${\bm T}_8$ commute with one another.  It is well known the eigenvalues  
 of these two matrices define the weights of the fundamental representation.
 The three eigenvectors of these two matrices can be denoted by $| \fracm 12 , \, 
 \fracm 1{2 {\sqrt 3}} > $,  $| - \fracm 12 , \,  \fracm 1{2 {\sqrt 3}} > $, and $| 0, \, - 
 \fracm 1{ {\sqrt 3}} > $ and satisfy
 \be  \eqalign{   {~~~~~~~~~~~}
 {\bm T}_3 \, | \fracm 12 , \,  \fracm 1{2 {\sqrt 3}} > ~&=~ \fracm 12 \,  \, | \fracm 12 , 
 \,  \fracm 1{2 {\sqrt 3}} >  ~~~~~~\,~~~, ~~~~~~  {\bm T}_8 \, | \fracm 12 , \,  \fracm 1{2 {\sqrt 3}} > 
 ~=~  \fracm 1{2 {\sqrt 3}} \, | \fracm 12 , \,  \fracm 1{2 {\sqrt 3}} >  ~~~~~\,~,\cr
  {\bm T}_3 \, | - \fracm 12 , \,  \fracm 1{2 {\sqrt 3}} > ~&=~ - \fracm 12 \,  \, | \fracm 12 , 
 \,  \fracm 1{2 {\sqrt 3}} >  ~~~~~~, ~~  {\bm T}_8 \, | -  \fracm 12 , \,  \fracm 1{2 {\sqrt 3}} > 
 ~=~  \fracm 1{2 {\sqrt 3}} \, |- \fracm 12 , \,  \fracm 1{2 {\sqrt 3}} >  ~~~,\cr
  {\bm T}_3 \, | 0 , \, - \fracm 1{ {\sqrt 3}} > ~&=~   0 
  ~~~~~~~~~~~~~~~~~~~~~~~~\,~, ~~~~~~\,  {\bm T}_8 \, | 0 , \,  \fracm 1{2 {\sqrt 3}} > 
 ~=~ -  \fracm 1{ {\sqrt 3}} \, | 0 , \,  - \fracm 1{ {\sqrt 3}} >  ~~~~~~,
} \ee
which is summarized as
\be
 {\bm T}_3 \, | t_1 , \,  t_2 > ~=~ t_1 \,  | t_1 , \,  t_2 >  ~~,~~
 {\bm T}_8 \, | t_1 , \,  t_2 > ~=~ t_2 \,  | t_1 , \,  t_2 >  ~~.
\ee
The remaining six su(3) generators can be arranged into ``raising operators'' and 
``lowering operators,''
 \be
 {\bm T}_{\pm} ~=~ {\bm T}_1~ \pm ~ i \, {\bm T}_2 ~~,~~
 {\bm U}_{\pm} ~=~ {\bm T}_6~ \pm ~  i \,{\bm T}_7 ~~,~~
 {\bm V}_{\pm} ~=~ {\bm T}_4~ \pm ~  i \, {\bm T}_5 ~~,
 \ee
whose effects are to generate `motions' between the eigenvectors.  We also have
\be
\left[ \,  {\bm T}_{+}  ~,~  {\bm T}_{-} \, \right] ~=~ 2 \,  {\bm T}_{3}
~~,~~ 
\left[ \,  {\bm U}_{+}  ~,~  {\bm U}_{-} \, \right] ~=~ {\sqrt 3}  \, {\bm T}_{8} ~-~   
{\bm T}_{3}  ~~,~~ 
\left[ \,  {\bm V}_{+}  ~,~  {\bm V}_{-} \, \right] ~=~ {\sqrt 3}  \, {\bm T}_{8} ~+~   
{\bm T}_{3} ~~,
\ee
along with a number of commutators which have the property that the matrices ${\bm 
T}_{3}$ and ${\bm T}_{8}$ do not appear on the rhs of them.  We also have the obvious 
results
\be  {
\left[ \,  {\bm T}_{\pm}  ~,~  {\bm T}_{\pm} \, \right] ~=~
\left[ \,  {\bm U}_{\pm}  ~,~  {\bm U}_{\pm} \, \right] ~=~
\left[ \,  {\bm V}_{\pm}  ~,~  {\bm V}_{\pm} \, \right] ~=~ 0 ~~.
} \label{Ro}
\ee

For the case of the one dimensional $N$ = 2 SUSY algebra one has two
``supercharges,'' ${\rm D}_1$ and ${\rm D}_2$, together with the time translation 
operator $\pa_{\t}$ satisfying the graded super Lie algebra bracket conditions
\be
\left\{  \, {\rm D}_{\rm I} ~,~ {\rm D}_{\rm J} \, \right\} ~=~ i 2 \d{}_{\rm I}
{}_{\rm J} \, \pa_{\t} ~~,~~ \left[  \, {\rm D}_{\rm I} ~,~ \pa_{\t} \, \right] ~=~ 0
~~,~~ \left[  \,  \pa_{\t} ~,~ \pa_{\t} \, \right] ~=~ 0 ~~.
\ee
The second of these relations show that $\pa_{\t}$ and ${\rm D}_1$ can be
chosen as members of a mutually commuting set.  It is here the usual 
Jordan-Chevalley decomposition appears to fail, as the only remaining generator 
${\rm D}_2$ cannot be used to create raising and lowering operators \cite{TH1}.

There is a much shorter argument that can be made as to why a representation
theory approach to space-time SUSY must begin in a different distinctive way.
The argument above shows how important is the role played by eigenvectors
(and their related space of eigenvalues) of a subset of the Lie algebra generators.  
Thus, any attempt to ``build'' a representation space approach for space-time 
SUSY based on the experience in Lie algebras would require the existence of
``eigenstates'' that can be acted upon by one or more ``supercharges.''  Since
supercharges map between the spaces of bosons and fermions (and vice-versa), 
clearly the existence of such ``eigenstates'' must be severely questioned, if not 
outright rejected.  The simplest route forward is to reject the existence of such
SUSY ``eigenstates.''

These observations lead to the idea that something new must be introduced 
in order to construct a representation theory for spacetime SUSY that has attributes
similar to those of the hugely successful representation theory of Lie algebras.
We have proposed that adinkras are the suitable candidates to study for this
purpose.

The key point of the Jordan-Chevalley decomposition used in Lie algebras is
that the maximal commuting matrices for the generators acting on any representation
yields a set of constants, the eigenvalues of the maximal commuting matrices, that are
intrinsic to each representation.  Guided by the study of adinkras, we have 
proposed in the context of the 1D, $N$-extended supersymmetry algebras,
there also exists a way to identify a similar set of constants for each supersymmetrical
representation.

On any of the valise adinkra, such as shown in Fig.\ \# 1
$$
\vCent
{\setlength{\unitlength}{1mm}
\begin{picture}(-20,-140)
\put(-49,0){\bf {{$\cal R\,=$} \# 1}}
\put(30,0){\bf {{$\cal R\,=$} \# 2}}
\put(-74,-36){\includegraphics[width=2.6in]{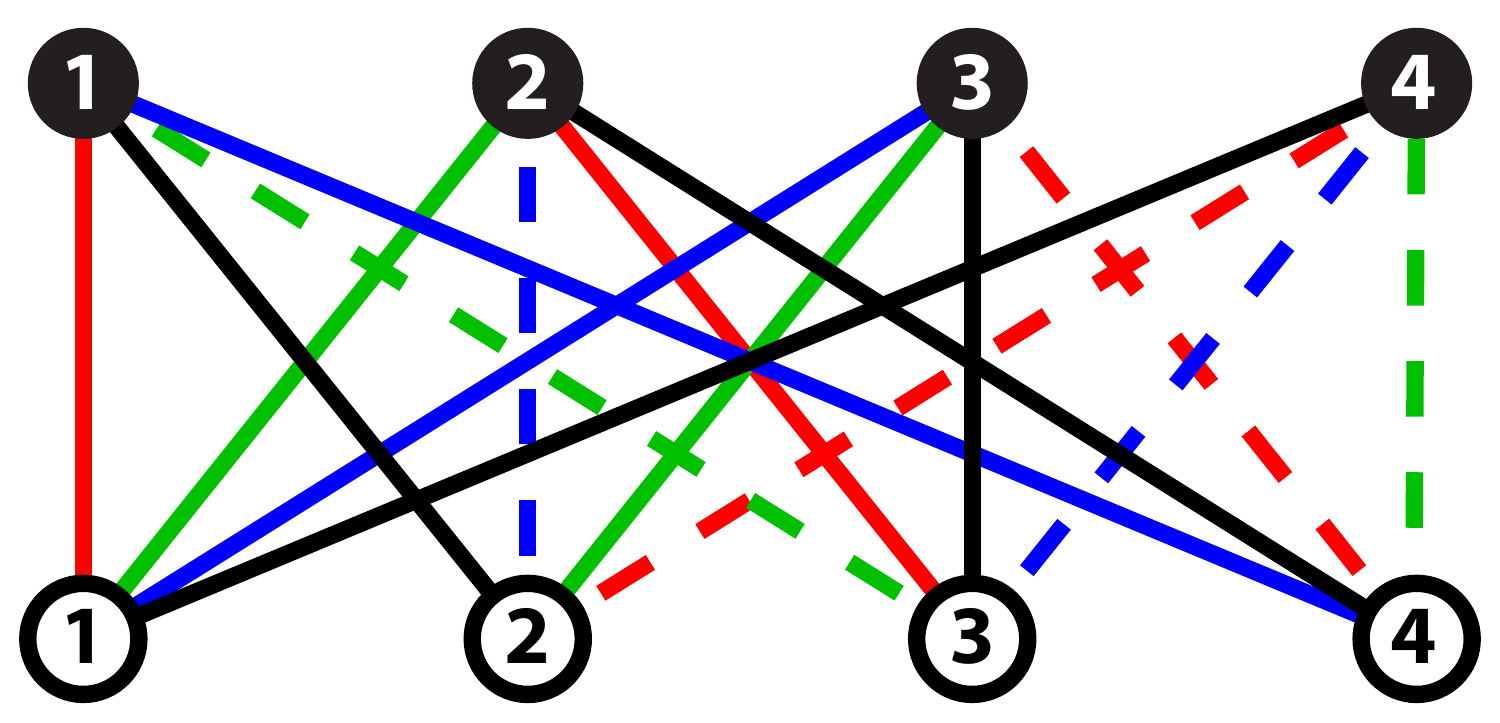}}
\put(6,-36.2){\includegraphics[width=2.6in]{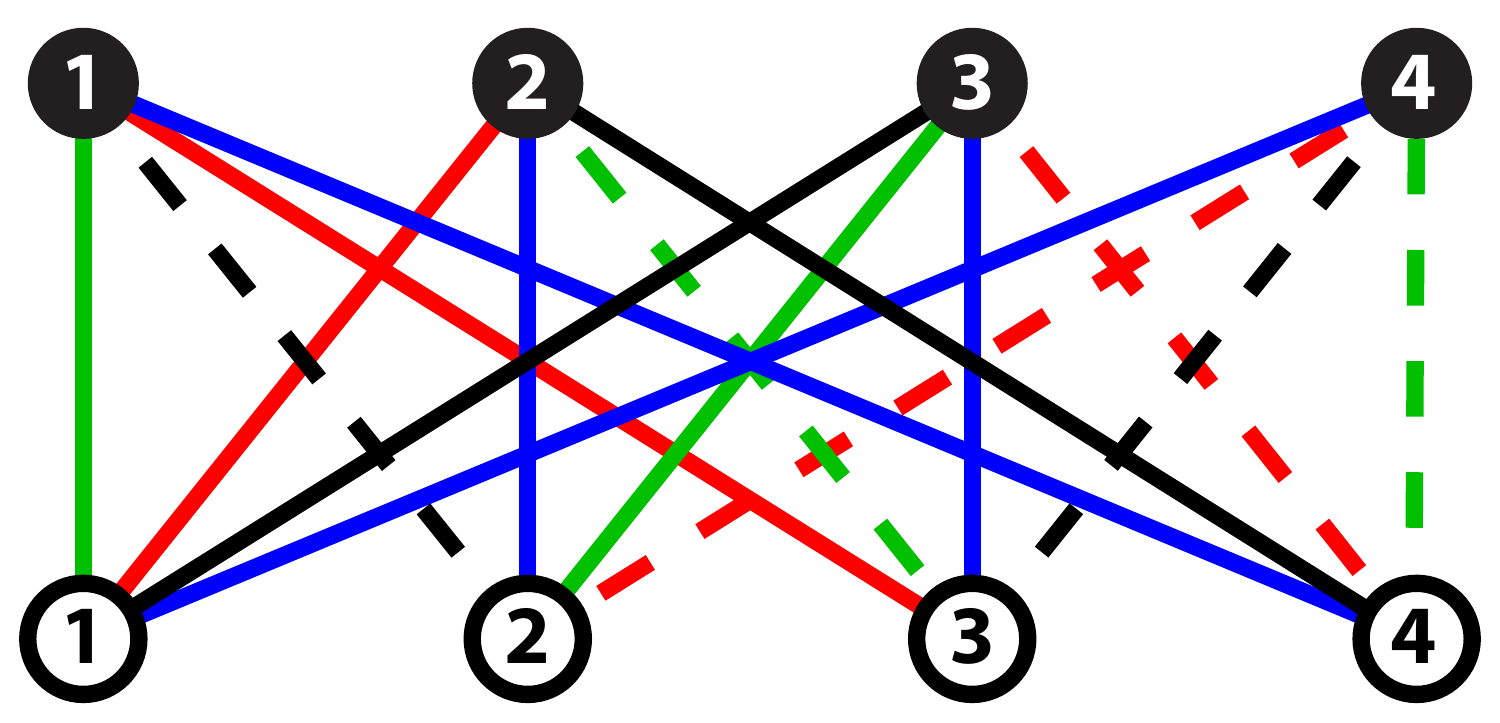}}
\put(-10,-44){\bf {{$\cal R\,=$} \# 3}}
\put(-34,-79){\includegraphics[width=2.6in]{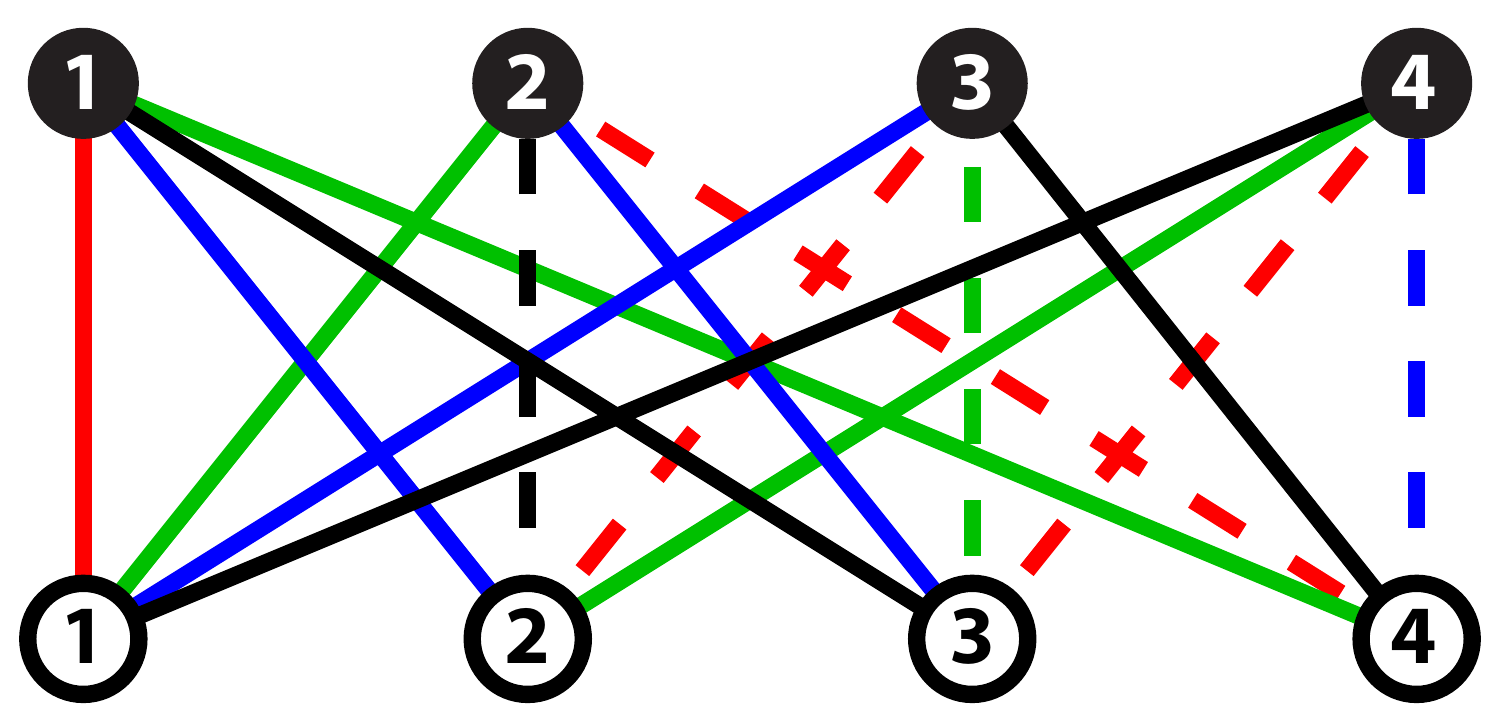}}
\put(-38,-86){\bf {Figure \# 1: Three valise adinkra graphs}}
\end{picture}}
$$
\vskip3.2in  \noindent
equations of the form
\be \eqalign{
{\rm D}{}_{{}_{\rm I}} \Phi_{i}^{(\cal R)} ~=~ i \, \left( {\rm L}{}_{{}_{\rm I}}^{(\cal R)} \right) 
{}_{i \, {\hat k}}  \, \, \Psi_{\hat k}^{(\cal R)}  ~~~,~~~
{\rm D}{}_{{}_{\rm I}} \Psi_{\hat k}^{(\cal R)} ~=~ \left( {\rm R}{}_{{}_{\rm I}}^{(\cal R)} \right)
{}_{{\hat k} \, i}  \, \pa_{\tau} \, \Phi_{i}^{(\cal R)}  ~~~,
}  \label{VH1}
\ee
are valid \cite{ENUF1,ENUF2}.  In the expression, $\Phi_{i}^{(\cal R)}$ denotes the i-th boson associated
with the $({\cal R})$-th adinkra and $\Psi_{\hat k}^{(\cal R)} $ denotes the ${\hat k}$-th 
fermion associated with the $({\cal R})$-th adinkra.  The constants $( {\rm L}{}_{{}_{\rm 
I}}^{(\cal R)}) {}_{i \, {\hat k}}$ and $( {\rm R}{}_{{}_{\rm I}}^{(\cal R)} ){}_{{\hat k} \, i} $
are the matrix elements of slightly modified versions of the adjacency matrices 
appropriate for the adinkra graphs.

It is possible to carry out the calculation of the commutator of two supercovariant
derivatives on all the fermions in any of the valises and due to dimensional analysis 
alone, the result of such a calculation must take the form \cite{adnkholor1,adnkholor2}
\be  \eqalign{
[\, {\rm D}{}_{{}_{\rm I}}  ~,~ {\rm D}{}_{{}_{\rm J}} \, ] \, \Psi_{\hat k}^{({\cal R})}  ~&=~ 
 2\, \left[ \,  {\Tilde V}^{({\cal R})}{}_{{}_{\rm I}}{}_{{}_{\rm J}} 
 \, \right]  {}_{{\hat k} \, 
{\hat \ell}}  \,  \pa_{\tau} \, \Psi_{\hat \ell}^{({\cal R})}     ~~~,
}  \label{VH2}
\ee
for some set of constants $[ \,  {\Tilde V}{}^{({\cal R})} {}_{{}_{\rm I}}{}_{{}_{\rm J}} \, ]
{}_{{\hat k} \, {\hat \ell}}$.  These constants are the components of the ``fermionic
holoraumy tensor'' for the $({\cal R})$-th adinkra.  Due to the results in (\ref{VH1})
we find
\be \eqalign{
\left[ \, {\Tilde V}{}_{\rI\rJ}^{({\cal R})}  \,\right] {}_\hi{\,}^\hk ~&=~ i\,  \frac12 \, 
\left[ \, (\,{\rm R}_\rI {}^{({\cal R})} \,)_\hi{}^j\>(\, {\rm L}_\rJ {}^{({\cal R})} \,
)_j{}^\hk ~-~ (\,{\rm R}_\rJ  {}^{({\cal R})} \, )_\hi{}^j\>(\,{\rm L}_\rI
{}^{({\cal R})} 
\,)_j{}^\hk  \, \right]
~~.
}  \label{VH4}
\ee
It is our contention that the data contained in $[ \,  {\Tilde V}{}^{({\cal R})} 
{}_{{}_{\rm I}}{}_{{}_{\rm J}} \, ] {}_{{\hat k} \, {\hat \ell}}$ should play the role 
for the classification of adinkras in the same way that the data contained in
the eigenvalues of the maximal commuting matrices plays for a Lie algebra.
However, the current paper is {\em {not}} about adinkras as the topics of 
study are 4D, $\cal N$ = 1 supermuliplets.  

The question we pursue here can be posed in the form below.

$~~~~$  $~~~~$  $~~~~$ $~~$ $~~~~$ {\it {If in compact Lie algebras 
it is the eigenvalues of a representation}} 
\newline $~~~~$  $~~~~$  $~~~~$ $~~$  $~~~~$ 
{\it {of the maximal commuting generators on a representation 
that provide}} 
\newline $~~~~$  $~~~~$  $~~~~$ $~~$  $~~~~$ 
{\it {the entrance to the development of a robust representation theory, and}} 
\newline $~~~~$  $~~~~$  $~~~~$ $~~$  $~~~~$ 
{\it {if the adinkra fermionic holoraumy tensor provides the entrance to the}} 
\newline $~~~~$  $~~~~$  $~~~~$ $~~$  $~~~~$ 
{\it {development of a robust adinkra representation theory, what calculable}} 
\newline $~~~~$  $~~~~$  $~~~~$ $~~$  $~~~~$ 
{\it {data, derivable from the supersymmetry transformation laws of a super-}} 
\newline $~~~~$  $~~~~$  $~~~~$ $~~$  $~~~~$ 
{\it {multiplet, play the analogous role for 4D, $\cal N$ = 1 supersymmetical theo-}} 
\newline $~~~~$  $~~~~$  $~~~~$ $~~$  $~~~~$ 
{\it {ries?}}

\noindent
By the end of this work, we hope to convince the reader there is a well
defined prescription by which this query may be answered.  Adinkras
will play the role of guide to this answer.

\newpage
\section{Evidence For A Weight Space Among All Minimal 4D, $\cal N$ = 1
Supermultiplets}
\label{s1a}

$~~~~$ It is well known the smallest representations of 4D, $\cal N$ = 1 
supersymmetry consist of four bosons and four fermions.  What is not
 generally recognized, however, is how numerous are such representations.  
 In addition to the well known chiral, vector, and tensor supermultiplets 
described below in the equations (\ref{QT1}), (\ref{QT2}), and (\ref{QT3}) 
\vskip 0.12in
$\bm {Chiral~Supermultiplet: ~(A, \, B, \,  \psi_a , \, F, \, G)}$
\be
 \eqalign{
{~~~~} {\rm D}_a A ~&=~ \psi_a  ~~~~~~~~~~~~~~,~~~~
{\rm D}_a B ~=~ i \, (\gamma^5){}_a{}^b \, \psi_b  ~~~~~~~~~, \cr
{\rm D}_a \psi_b ~&=~ i\, (\gamma^\mu){}_{a \,b}\,  \partial_\mu A 
~-~  (\gamma^5\gamma^\mu){}_{a \,b} \, \partial_\mu B ~-~ i \, C_{a\, b} 
\,F  ~+~  (\gamma^5){}_{ a \, b} G  ~~, \cr
{\rm D}_a F ~&=~  (\gamma^\mu){}_a{}^b \, \partial_\mu \, \psi_b   
~~~,~~~ 
{\rm D}_a G ~=~ i \,(\gamma^5\gamma^\mu){}_a{}^b \, \partial_\mu \,  
\psi_b  ~~~,
} \label{QT1}
\ee \indent
$\bm {Vector~Supermultiplet:~ (A{}_{\mu} , \, \l_b , \,  {\rm d})}$
\be
\eqalign{
{~~~~} {\rm D}_a \, A{}_{\mu} ~&=~  (\gamma_\mu){}_a {}^b \,  \l_b  ~~~, \cr
{\rm D}_a \l_b ~&=~   - \,i \, \fracm 14 ( [\, \gamma^{\mu}\, , \,  \gamma^{\nu} 
\,]){}_a{}_b \, (\,  \partial_\mu  \, A{}_{\nu}    ~-~  \partial_\nu \, A{}_{\mu}  \, )
~+~  (\gamma^5){}_{a \,b} \,    {\rm d} ~~,  {~~~~~~~}  \cr
{\rm D}_a \, {\rm d} ~&=~  i \, (\gamma^5\gamma^\mu){}_a {}^b \, 
\,  \partial_\mu \l_b  ~~~, \cr
}  \label{QT2}
\ee
 \indent
$\bm {Tensor~Supermultiplet: ~(\varphi, \, B{}_{\mu \, \nu }, \,  \chi_a )}$
\be
 \eqalign{
{\rm D}_a \varphi ~&=~ \chi_a  ~~~,~~~~~ 
{\rm D}_a B{}_{\mu \, \nu } ~=~ -\, \fracm 14 ( [\, \gamma_{\mu}
\, , \,  \gamma_{\nu} \,]){}_a{}^b \, \chi_b  ~~~, \cr
{\rm D}_a \chi_b ~&=~ i\, (\gamma^\mu){}_{a \,b}\,  \partial_\mu \varphi 
~-~  (\gamma^5\gamma^\mu){}_{a \,b} \, \e{}_{\mu}{}^{\r \, \s \, \t}
\partial_\r B {}_{\s \, \t}~~, {~~~~~~~~~~~~~~\,~~}
}  \label{QT3}
\ee
as well as the parity opposites to the vector and tensor supermultiplets
(i.\ e.\ the axial vector and axial tensor supermultiplets) given in 
(\ref{QT2a}) and (\ref{QT3a}) \vskip 0.12in
$\bm {Axial-Vector~Supermultiplet:~ (U{}_{\mu} , \, {\Tilde \l}_b , \,  {\Tilde {\rm d}}
)}$
\be
\eqalign{
{~~~~} {\rm D}_a \, U{}_{\mu} ~&=~ i\, (\gamma^5 \gamma_\mu){}_a {}^b 
\,  {\Tilde \l}_b  ~~~, \cr
{\rm D}_a {\Tilde \l}_b ~&=~  \, \fracm 14 ( \gamma^5 [\, \gamma^{\mu}\, , \,  
\gamma^{\nu} \,]){}_a{}_b \, (\,  \partial_\mu  \, U{}_{\nu}    ~-~  \partial_\nu 
\, U{}_{\mu}  \, ) ~+~ i \, C{}_{a \,b} \, {\Tilde {\rm d}} ~~,  {~~~~~~~}  \cr
{\rm D}_a \,  {\Tilde {\rm d}} ~&=~  - \, (\gamma^\mu){}_a {}^b \, 
\,  \partial_\mu {\Tilde \l}_b  ~~~, 
\cr}  \label{QT2a}
\ee
 \indent
$\bm {Axial-Tensor~Supermultiplet: ~( {\Tilde {\varphi}}, \, 
C{}_{\mu \, \nu }, \,  {\Tilde {\chi}} {}_a )}$
\be
 \eqalign{
{\rm D}_a {\Tilde {\varphi}} ~&=~ - i \, (\gamma^\mu){}_{a}{}^{b} \, {\Tilde {\chi}}{}_a  ~~~,~~~~~ 
{\rm D}_a C{}_{\mu \, \nu } ~=~ i\, \fracm 14  ( \gamma^5 [\, \gamma_{\mu}
\, , \,  \gamma_{\nu} \,]){}_a{}^b \,  {\Tilde {\chi}}{}_b  ~~~, \cr
{\rm D}_a  {\Tilde {\chi}}{}_b ~&=~ -\, (\gamma^5 \gamma^\mu){}_{a \,b}\,  \partial_\mu 
 {\Tilde {\varphi}}
~-~  i \, (\gamma^\mu){}_{a \,b} \, \e{}_{\mu}{}^{\r \, \s \, \t}
\partial_\r C{}_{\s \, \t}~~, {~~~~~~~~~~~~~~\,~~}
}  \label{QT3a}
\ee
there are an equal number of supermultiplets that are obtained via Hodge
duality to the chiral and vector supermultiplets above.

A Hodge dual version of a supermultiplet occurs when the field strength for a
field of a given Lorentz representation is replaced by the Hodge dual field
strength of a distinct field.  The tensor supermultiplet \cite{TnsR} may be 
regarded as a Hodge dual formulation of the usual chiral supermultiplet.  One 
can take the derivative of one of propagating spin-0 fields (which can be 
regarded as its field strength) and replace it by the field strength of a two-form.
Both theories are off-shell.  If one attempts to perform this process on
both propagating spin-0 fields \cite{FrdMn}, one obtains a supermultiplet
that does {\em {not}} possess a closed off-shell supersymmetry algebra.

However, such Hodge duality replacement need not be restricted to
propagating spin-0 fields, as it has long been known \cite{CLShist3} that auxiliary
spin-0 fields can be subjected to the same processes.  As the chiral
supermultiplet has two auxiliary spin-0 fields, this implies there are
{\em {three}} Hodge dual variants:

(a.) one arises from the replacement of the scalar auxiliary field by
a gauge \newline $~~~~~~~~~~~$ 3-form,

(b.) one arises from the replacement of the pseudoscalar auxiliary field by
 \newline $~~~~~~~~~~~$ a gauge 3-form, and

(d.) one arises from the replacement of {\em {both}} the scalar and
pseudoscalar  \newline $~~~~~~~~~~~$ auxiliary fields by  gauge 3-forms.
\vskip0.05in  \noindent
These are shown in equations (\ref{QTd1}), (\ref{QTd2}), and
(\ref{QTd3}), respectively. 

$\bm {Hodge-Dual~ \#1~Chiral~Supermultiplet: ~(A, \, B, \,  \psi_a , \, {\rm f}_{\mu
 \, \nu \, \rho}, \, G)}$
\be
 \eqalign{
{~~~~} {\rm D}_a A ~&=~ \psi_a  ~~~~~~~~~~~~~~,~~~~
{\rm D}_a B ~=~ i \, (\gamma^5){}_a{}^b \, \psi_b  ~~~~~~~~~, \cr
{\rm D}_a \psi_b ~&=~ i\, (\gamma^\mu){}_{a \,b}\,  \partial_\mu A 
~-~  (\gamma^5\gamma^\mu){}_{a \,b} \, \partial_\mu B ~-~ i \,  \frac 1{3!} \, C_{a\, b} 
\, (\epsilon{}^{\s}{}^{\mu}{}^{\nu}{}^{\rho} \, \pa_{\s} {\rm f}_{\mu
 \, \nu \, \rho})  ~+~  (\gamma^5){}_{ a \, b} G  ~~, \cr
{\rm D}_a {\rm f}_{\mu \, \nu \, \rho} ~&=~ -\,  (\gamma^\s){}_a{}^b \,  
\epsilon{}_{\s}{}_{\mu}{}_{\nu}{}_{\rho} \, \psi_b   
~~~,~~~ 
{\rm D}_a G ~=~ i \,(\gamma^5\gamma^\mu){}_a{}^b \, \partial_\mu \,  
\psi_b  ~~~,} \label{QTd1}
\ee 
$\bm {Hodge-Dual~ \#2~Chiral~Supermultiplet: ~(A, \, B, \,  \psi_a , \, F, \, {\rm 
g}_{\mu \, \nu \, \rho})}$
\be
 \eqalign{
{~~~~} {\rm D}_a A ~&=~ \psi_a  ~~~~~~~~~~~~~~,~~~~
{\rm D}_a B ~=~ i \, (\gamma^5){}_a{}^b \, \psi_b  ~~~~~~~~~, \cr
{\rm D}_a \psi_b ~&=~ i\, (\gamma^\mu){}_{a \,b}\,  \partial_\mu A 
~-~  (\gamma^5\gamma^\mu){}_{a \,b} \, \partial_\mu B ~-~ i \, C_{a\, b} 
\,F  ~+~  \frac 1{3!} \, (\gamma^5){}_{ a \, b}
\, (\epsilon{}^{\s}{}^{\mu}{}^{\nu}{}^{\rho} \, \pa_{\s} {\rm g}_{\mu
 \, \nu \, \rho})  ~~, \cr
{\rm D}_a F ~&=~  (\gamma^\mu){}_a{}^b \, \partial_\mu \, \psi_b   
~~~,~~~ 
{\rm D}_a {\rm g}_{\mu \, \nu \, \rho} ~=~ -\,  (\gamma^5 \gamma^\s){}_a{}^b \,  
\epsilon{}_{\s}{}_{\mu}{}_{\nu}{}_{\rho} \, \psi_b  
 ~~~,
} \label{QTd2}
\ee \indent
$\bm {Hodge-Dual~ \#3~Chiral~Supermultiplet: ~(A, \, B, \,  \psi_a , \, 
{\rm f}_{\mu \, \nu \, \rho}, \, {\rm g}_{\mu \, \nu \, \rho})}$
\be
 \eqalign{
{~~~~} {\rm D}_a A ~&=~ \psi_a  ~~~~~~~~~~~~~~,~~~~
{\rm D}_a B ~=~ i \, (\gamma^5){}_a{}^b \, \psi_b  ~~~~~~~~~, \cr
{\rm D}_a \psi_b ~&=~ i\, (\gamma^\mu){}_{a \,b}\,  \partial_\mu A 
~-~  (\gamma^5\gamma^\mu){}_{a \,b} \, \partial_\mu B \cr
&~~~~-~ i \,  
\frac 1{3!} \, C_{a\, b} \, (\epsilon{}^{\s}{}^{\mu}{}^{\nu}{}^{\rho} \, 
\pa_{\s} {\rm f}_{\mu \, \nu \, \rho})
 ~+~  \frac 1{3!} \, (\gamma^5){}_{ a \, b}
\, (\epsilon{}^{\s}{}^{\mu}{}^{\nu}{}^{\rho} \, \pa_{\s} {\rm g}_{\mu
 \, \nu \, \rho})   ~~, \cr
{\rm D}_a {\rm f}_{\mu \, \nu \, \rho} ~&=~ -\,  (\gamma^\s){}_a{}^b \,  
\epsilon{}_{\s}{}_{\mu}{}_{\nu}{}_{\rho} \, \psi_b    
~~~,~~~ 
{\rm D}_a {\rm g}_{\mu \, \nu \, \rho} ~=~ -\,  (\gamma^5 \gamma^\s){}_a{}^b \,  
\epsilon{}_{\s}{}_{\mu}{}_{\nu}{}_{\rho} \, \psi_b    ~~~.
} \label{QTd3}
\ee 

The process can also be applied to the auxiliary spin-0 fields of the vector
and axial-vector supermultiplets. \newline
\indent
$\bm {Hodge-Dual~Vector~Supermultiplet:~ (A{}_{\mu} , \, \l_b , \,  {\rm d}
{}_{\mu \, \nu \, \rho} 
)}$
\be
\eqalign{
{~~~~~~~~~~~~~~~~} {\rm D}_a \, A{}_{\mu} ~&=~  (\gamma_\mu){}_a 
{}^b \,  \l_b  ~~~, \cr
{\rm D}_a \l_b ~&=~   - \,i \, \fracm 14 ( [\, \gamma^{\mu}\, , \, \gamma^{\nu} 
\,]){}_a{}_b \, (\,  \partial_\mu  \, A{}_{\nu}    ~-~  \partial_\nu \, A{}_{\mu}  \, )
~+~  \frac 1{3!} \, (\gamma^5){}_{ a \, b} \, (\epsilon{}^{\s}{}^{\mu}{}^{\nu}
{}^{\rho} \, \pa_{\s} {\rm d}{}_{\mu \, \nu \, \rho} )  ~~,  {~~~~~~~}  \cr
{\rm D}_a \, {\rm d} ~&=~  i \, (\gamma^5\gamma^\mu){}_a {}^b \, 
\,  \partial_\mu \l_b  ~~~, \cr
}  \label{QTd2a}
\ee \indent
$\bm {Hodge-Dual~ Axial-Vector~Supermultiplet:~ (A{}_{\mu} , \, {\Tilde \l}
{}_b , \,  {\Tilde {\rm d}}{}_{\mu \, \nu \, \rho} )}$
\be
\eqalign{
{~~~~} {\rm D}_a \, U{}_{\mu} ~&=~ i\, (\gamma^5 \gamma_\mu){}_a {}^b 
\,  {\Tilde \l}_b  ~~~, \cr
{\rm D}_a {\Tilde \l}_b ~&=~  \, \fracm 14 ( \gamma^5 [\, \gamma^{\mu}\, , \,  
\gamma^{\nu} \,]){}_a{}_b \, (\,  \partial_\mu  \, U{}_{\nu}    ~-~  \partial_\nu 
\, U{}_{\mu}  \, ) ~+~ i \,  \frac 1{3!} \, C{}_{a \,b} \, \, (\epsilon{}^{\s}{}^{\mu}
{}^{\nu}{}^{\rho} \, \pa_{\s} {\Tilde {\rm d}}{}_{\mu \, \nu \, \rho} )  ~~,    \cr
{\rm D}_a \,  {\Tilde {\rm d}} ~&=~  - \, (\gamma^\mu){}_a {}^b \, \partial_\mu 
 {\Tilde \l}_b    ~~~, \cr
}  \label{QTd2b}
\ee
So all told, there are ten minimal {\em{off-shell}} 4D, $\cal N$ = 1 supermultiplets.

In the work of \cite{HoLoRmY4D}, the concept of ``4D, $\cal N$ = 1 Lorentz
fermionic holoraumy'' was introduced.  The holoraumy of a supermultiplet is
found from the {\em {commutator}} of two superspace fermionic derivatives
(i.e.\ the commutator of two supercharges) on the fermionic fields in
the supermultiplet.  Executing this calculation on the chiral, vector, and 
tensor supermultplets, respectively, yields the results in (\ref{HCSf}), 
(\ref{HVSf}), and (\ref{HTSf}).  We may assign a representation space
label $(\widehat {\cal R})$ to each supermultiplet such that 
\vskip5pt
\begin{table}[h]
\begin{center}
\footnotesize
\begin{tabular}{|c|c|}\hline 
Supermultiplet  & Supermultiplet Rep Label $(\widehat {\cal R})$ \\ \hline \hline
Chiral Supermultiplet  & (CS) \\ \hline
Vector Supermultiplet  & (VS) \\ \hline
Tensor Supermultiplet  & (TS) \\ \hline
\end{tabular}
\vskip15pt {{{\bf {Table}} {\bf {1:}}
Representation Label For Supermultiplets}}
\end{center}
\end{table}
\noindent
which allows us to collectively write the results of the calculations

$\bm {Chiral~Supermultiplet~Fermion}$
\be   \eqalign{
[\, {\rm D}{}_{a} ~,~ {\rm D}{}_{b} \, ] \,\psi_{c} ~&=~   \left[{ {\bm H}}{}^{\mu}
{}^{(CS)}\right]{}_{a \, b \, c}{}^d \, \left(\pa_{\mu} \psi_{d}  \, \right)  ~~~, 
} \label{HCSf} 
\ee

$\bm {Vector~Supermultiplet~Fermion}$
\be \eqalign{
[\, {\rm D}{}_{a} ~,~ {\rm D}{}_{b} \, ] \,\lambda_{c} ~&=~  \left[{ {\bm H}}
{}^{\mu}{}^{(VS)}\right]{}_{a \, b \, c}{}^d \, \left(\pa_{\mu} \l_{d}  \, \right)   ~~~,
} \label{HVSf} 
\ee

$\bm {Tensor~Supermultiplet~Fermion}$
\be \eqalign{
[\, {\rm D}{}_{a} ~,~ {\rm D}{}_{b} \, ] \,\chi_{c} ~&=~  \left[{ {\bm H}}{}^{\mu}
{}^{(TS)}\right]{}_{a \, b \, c}{}^d \, \left(\pa_{\mu} \chi_{d}  \, \right)  ~~~,
}  \label{HTSf}
\ee
 in a
concise form by the introduction of the notational device of the 4D ``fermionic holoraumy
tensors''
\be  \eqalign{ {~~~~~~~~~~~}
\left[{ {\bm H}}{}^{\mu}( {\rm p}_{({\cal R})},  {\rm q}_{({\cal R})}, {\rm r}_{({\cal R})},   
{\rm s}_{({\cal R})}  \,) \right]{}_{a \, b \, c}{}^d  ~&=~ -i 2 \, {\Big [} \, \,  {\rm p}_{({\cal 
R})} \, C_{ab} \, (\gamma^{\m})_c{}^{d}  ~+~   {\rm q}_{({\cal R})} \, (\gamma^{5}
)_{ab} (\gamma^{5}\gamma^{\m})_{c}{}^{d}     \cr 
&~~~~~~~~~~~~~
 ~+~ {\rm r}_{({\cal R})} \,  (\gamma^5 \gamma^{\mu} )_{ab} \, 
 (\gamma^5)_{c}{}^{d}     \cr
 &~~~~~~~~~~~~~~+~  \fracm 12 \,  {\rm s}_{({\cal R})} \, (\gamma^5 \gamma^{\nu
 } )_{ab} \, (\gamma^5 \, [ \gamma {}_{\nu}\,, \, \gamma^{\mu} ])_{c}{}^{d}    \, \, {\Big ]}    
~~~,
}   \label{4DHs}
\ee
where the quantities ${\rm p}_{({\cal R})}$, ${\rm q}_{({\cal R})}$, ${\rm r}_{({\cal R})}$, 
and ${\rm s}_{({\cal R})}$ are integers.  Thus we find from explicit calculation for each 
of these three supermultiplets the results shown in Table \# 2.
\vskip5pt
\begin{table}[h]
\begin{center}
\footnotesize
\begin{tabular}{|c|c|c|c|c|}\hline 
$(\widehat {\cal R})$  & ${\rm p}_{({\cal R})}$ & ${\rm q}_{({\cal R})}$ &
${\rm r}_{({\cal R})}$ & ${\rm s}_{({\cal R})}$ \\ \hline \hline
(CS) & 0 & 0 & 0 & 1  \\ \hline
(VS) & 1 & 1 & 1 & 0  \\ \hline
(TS) & - 1 & 1 & - 1 & 0   \\ \hline
\end{tabular}
\vskip15pt {{{\bf {Table}} {\bf {2:}}
Holoraumy Integers For (CS), (VS), and (TS) Supermultiplets}}
\end{center}
\end{table} \noindent
The holoraumy for the parity-swapped version of each supermultiplet is found
from
\be   \eqalign{
\left[{ {\bm H}}{}^{\mu}{}^{(AVS )}\right]{}_{a \, b \, c}{}^d \, ~&=~ 
(\gamma^{5} )_{c}{}^{e} \, \left[{ {\bm H}}{}^{\mu}{}^{(VS)}\right]
{}_{a \, b \, e}{}^f ( \gamma^{5} )_{f}{}^{d} \,  ~~,  ~~
\left[{ {\bm H}}{}^{\mu}{}^{(ATS)}\right]{}_{a \, b \, c}{}^d \, ~=~ 
(\gamma^{5} )_{c}{}^{e} \, \left[{ {\bm H}}{}^{\mu}{}^{(TS)}\right]
{}_{a \, b \, e}{}^f ( \gamma^{5} )_{f}{}^{d} \,  ~~~~,
}  \label{HTTswap1}
\ee
which leads to the results shown in Table 3 where (AVS) and (ATS) are the 
representation labels for the axial vector and axial tensor supermultiplets, 
respectively.

\vskip5pt
\begin{table}[h]
\begin{center}
\footnotesize
\begin{tabular}{|c|c|c|c|c|}\hline 
$(\widehat {\cal R})$  & ${\rm p}_{({\cal R})}$ & ${\rm q}_{({\cal R})}$ &
${\rm r}_{({\cal R})}$ & ${\rm s}_{({\cal R})}$ \\ \hline \hline
(AVS) & -1 & -1 & 1 & 0  \\ \hline
(ATS) &  1 & - 1 & - 1 & 0   \\ \hline
\end{tabular}
\vskip15pt {{{\bf {Table}} {\bf {3:}}
Holoraumy Integers For (AVS), and (ATS) Supermultiplets}}
\end{center}
\end{table} 
Regarding the entries in each row as the components of vectors and using the
inner product defined by
\be
 \eqalign{ {~~~~~~~}
{\widehat  {\cal G}} [  ({\widehat {\cal R}}) , ({\widehat {\cal R}}^{\prime}) ]
~&=~ \frc 13 \,  {\Big [}   ~  {\rm p}_{({\cal R})} \,   {\rm p}_{({\cal R}^{\prime})} 
~+~    {\rm q}_{({\cal R})} \,   {\rm q}_{({\cal R}^{\prime})}  ~+~ {\rm r}_{({\cal R})} 
\, {\rm r}_{({\cal R}^{\prime})} ~+~  3\,  {\rm s}_{({\cal R})} \,   {\rm s}_{({\cal 
R}^{\prime})}  ~ {\Big ]}    \cr
~&=~  \frc 1{1,536} \, {\Big \{ } \,\, 
 (\gamma^{\a})_c^{~e}  \,  [{ {\bm H}}{}^{\mu}{
}^{({\widehat {\cal R}})} ]{}_{a \, b \, e}{}^f  \, (\gamma_{\a})_f^{
~d} \, [ { {\bm H}}{}_{\mu}{}^{({\widehat {\cal R}}^{\prime})} ]{}^{a \, 
b}{}_{d}{}^c                \cr
&~~~~~~~~~~~~~~~~~~~
-~ \, \frc 18   
  ([\, \gamma^{\a}  ~,~ \gamma^{\b} \,])_c^{~e} \,  
[{ {\bm H}}{}^{\mu}{}^{({\widehat {\cal R}})} ]{}_{a \, b \, 
e}{}^f  \, ([\, \gamma_{\a}  ~,~ \gamma_{\b} \,])_f^{~d} \, [ { {\bm 
H}}{}_{\mu}{}^{({\widehat {\cal R}}^{\prime})} ]{}^{a \, b}{}_{d}{}^c
  ~ {\Big \} }     \cr
~&=~ -\, \frc 1{1,536} \, {\Big \{ } \,\, [{ {\bm H}}{}^{\mu}{}^{({\widehat {\cal R}})}
]{}_{a \, b \, c}{}^d  \, [ { {\bm H}}{}_{\mu}{}^{({\widehat {\cal 
R}}^{\prime})} ]{}^{a \, b}{}_{d}{}^c   \cr
&~~~~~~~~~~~~~~~~~~~
-~ \,    (\gamma^5 \gamma^{\a})_c^{~e}  \,  [{ {\bm H}}{}^{\mu}{}^{({\widehat 
{\cal R}})} ]{}_{a \, b \, e}{}^f  \, (\gamma^5 \gamma_{\a})_f^{~d} \, [ { {\bm 
H}}{}_{\mu}{}^{({\widehat {\cal R}}^{\prime})} ]{}^{a \, b}{}_{d}{}^c    \cr
&~~~~~~~~~~~~~~~~~~
~+~ \,    (\gamma^5)_c^{~e}  \,  [{ {\bm H}}{}^{\mu}{}^{({\widehat {\cal R}})} 
]{}_{a \, b \, e}{}^f  \, (\gamma^5)_f^{~d} \, [ { {\bm H}}{}_{\mu}{}^{({\widehat 
{\cal R}}^{\prime})} ]{}^{a \, b}{}_{d}{}^c  ~ {\Big \} }  ~~~,  {~~~~~~~~~~~~~}
}  \label{GdGET2}
\ee
we obtain
\be
{\widehat  {\cal G}} [  ({\widehat {\cal R}}) , ({\widehat {\cal R}}^{\prime}) ] 
~=~ \left[\begin{array}{ccccc}
~1 & ~0 &  ~0  &~0 &  ~0 \\
~0 & ~~1 &  -\, \fracm 13 & -\, \fracm 13 &  -\, \fracm 13\\
~0 &  -\, \fracm 13 &  ~~ 1 & ~ -\, \fracm 13 &  -\, \fracm 13 \\
~0 &  -\, \fracm 13 &  -\, \fracm 13 & ~1 &  ~ -\, \fracm 13 \\
~0 &  -\, \fracm 13 &  -\, \fracm 13 & ~ -\, \fracm 13 &  ~1 \\
\end{array}\right]      ~~~,
\label{Gdgt3b}
\ee
where the row and column labels run over (CS), (VS), (TS), (AVS), and
(ATS), respectively.

These results have a simple geometrical interpretation.  The (CS)
representation lies in a spatial direction that is orthogonal to that of
the other representations.  Taking a three dimensional projection of
the remaining supermultiplet representations yields the image in Fig.\# 2.
\begin{figure}[ht]
\begin{center}
\begin{picture}(70,26)
\put(08,-33){\includegraphics[height =2.2in]{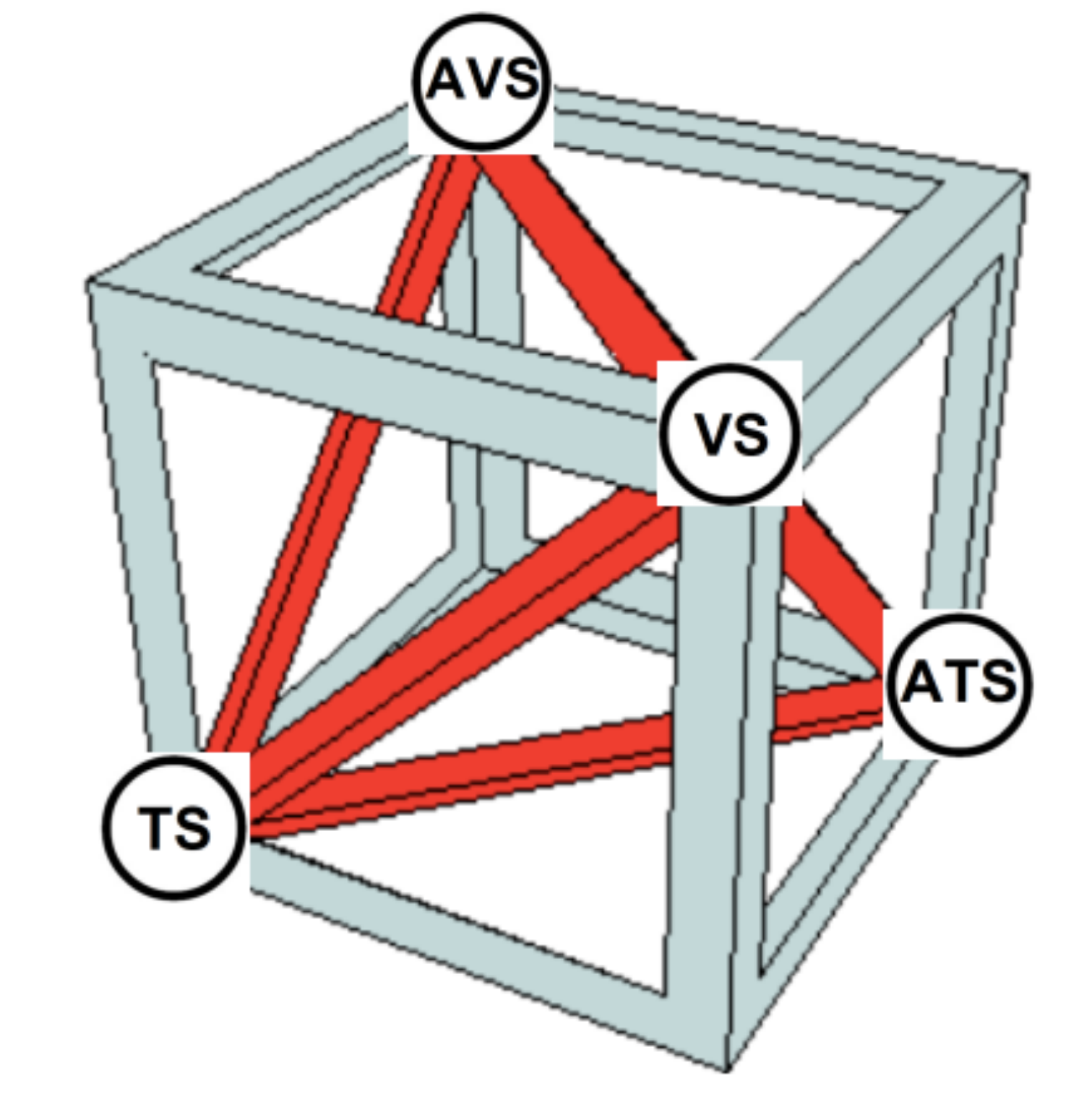}}
\put(-32,-40){{{\bf {Figure}} {\bf {2:}}
Cubically Inscribed Tetrahedron With Supermultiplets At  }}
\put(16,-45){{
Intersecting Vertices}}
\end{picture}
\end{center}
\label{TetMin}
\end{figure}
 \vskip1.2in
 \begin{center}
$~~~$
\end{center}
This structure begs for a simple explanation and it is easy to find one.

Let us replace the supermultiplets represented by the labeled balls, 
with unit vectors that point from the origin placed at the center of the 
cube to the vertices previously occupied by the balls.  This leads to 
the image in Fig. \#3. 
\begin{figure}[ht]
\begin{center}
\begin{picture}(70,26)
\put(08,-33){\includegraphics[height =2.2in]{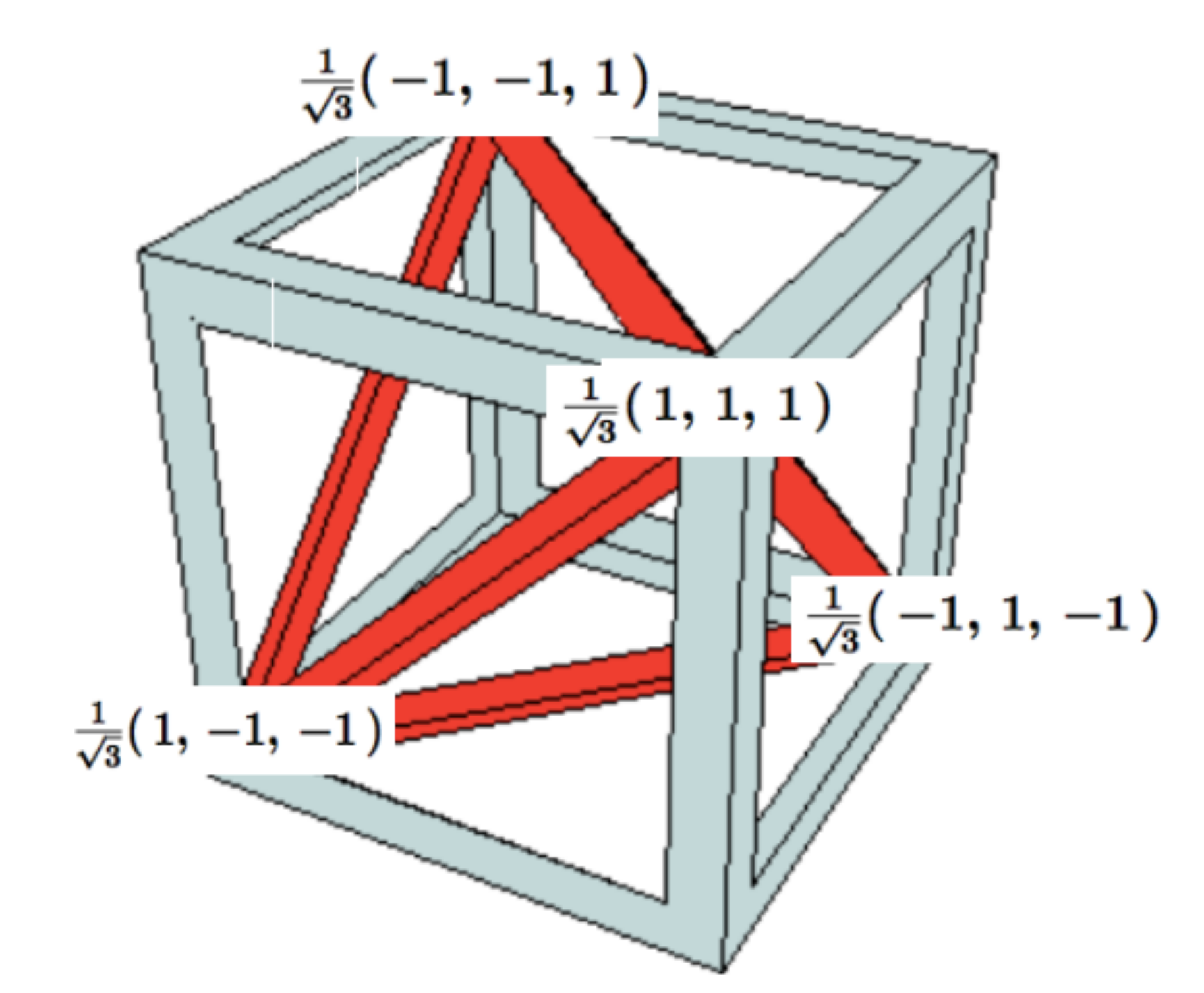}}
\put(-42,-40){{{\bf {Figure}} {\bf {3:}}
Coordinates Of Tetrahedron With Supermultiplets At Vertices }}
\end{picture}
\end{center}
\label{TetMin2}
\end{figure}
 \vskip1.0in
 \begin{center}
$~~~$
\end{center}
We know every vertex of the cube can be associated with one of the eight vectors
$\fracm 1{\sqrt 3}$ ($\pm 1$, $\pm 1$, $\pm 1$).  As such and seen in Fig.\ \# 3
 it is obvious that the common feature shared by the vertices occupied by
supermultiplets is these all possess even numbers of minus signs.

We can also ask about the representations that are obtained by Hodge dualities
from the chiral and vector supermultiplets\footnote{The tensor and axial-tensor
supermultiplets have no off-shell Hodge dual representations.}.

All of the supermultiplets that are Hodge dual to the chiral supermultiplet have
the same values of ${\rm p}_{({\cal R})}$, ${\rm q}_{({\cal R})}$, ${\rm r}_{({\cal R})}$, 
and ${\rm s}_{({\cal R})}$ as those that specify the chiral supermultiplet.  In a
similar manner, the Hodge duals to the vector and axial vector supermultiplets
have the same values of ${\rm p}_{({\cal R})}$, ${\rm q}_{({\cal R})}$, ${\rm r}_{({
\cal R})}$, and ${\rm s}_{({\cal R})}$ as those of the vector and axial vector supermultiplets.
So the vertices in Fig.\ \# 3 located at $\fracm 1 {\sqrt 3} (1, \, 1, \, 1)$ and $\fracm 1 
{\sqrt 3}  (-1, \, -1, \, 1)$ are doubly occupied when the Hodge dual
supermultiplets are included.

Also we note these degeneracies can be lifted by including another intrinsic 
property of adinkras.   Namely, the nodes that appear in a general adinkra can 
appear at more than the two heights that occur for valise adinkras.  By use of 
integrals or derivatives, the height of nodes may be changed.  When this property 
is taken into account, all the degeneracies seen in this chapter disappear.

In the next chapter, we wish to go beyond the minimal representations of
4D, $\cal N$ = 1 SUSY in this line of reasoning and explore how the concept
of fermionic holoraumy makes its appearance larger representations of
4D $\cal N$ = 1 supersymmetry.

\newpage
\section{The Four Dimensional CLS}
\label{s2a}

The complex linear supermultiplet \cite{CLShist1,CLShist2,CLShist3} 
propagates spin-0 degrees of freedom described by a scalar $\bs K$ and
a psuedoscalar $\bs L$ along with spin-1/2 degrees of freedom described 
by a Majorana spinor $ {\bs \zeta}_a$  as described by a Lagrangian 
$\mathcal{L}_{\text{\CLS}}$ whose explicit form is given in the form,
\be
\eqalign{  
\mathcal{L}_{\text{\CLS}} &= -\frc{1}2\partial_\mu \bs K \partial^\mu \bs K ~-~ \frc{1}2 
\partial_\mu \bs L \partial^\mu \bs L ~-~ \frc{1}2 \bs M^2 ~-~ \frc{1}2 \bs N^2 ~+~
\frc{1}4\, {\bs U}_\mu \, {\bs U}^\mu  ~+~ \frc{1}4 \, {\bs V}_\mu \, {\bs V}^\mu \cr
&~~~\, +~\frc{1}2i (\gamma^\mu)^{ab} \, {\bs \zeta}_a \partial_\mu {\bs \zeta}_b  
~+~ i\, {\bs \rho}_a C^{ab} \, {\bs\beta}_b ~~~,
}  \label{CLSact} \ee
and is invariant (up to total derivatives) under the transformation laws
\be
\eqalign{
\rD_a\,\bK &= \Br_a - \Bz_a ~~~, \cr
\rD_a\,\bL &= i(\gamma^5)_a^{~b} (\Br_b + \Bz_b) ~~~, \cr
\rD_a\,\bM &= \BB_a -\frac{1}{2} (\gamma^\mu)_a^{~b}\,(\partial_\mu 
\Br_b)  ~~~, \cr
\rD_a\,\bN &=  -i (\gamma^5)_a^{~b}\,\BB_b +\frac{i}{2} (\gamma^5 
\gamma^\mu)_a^{~b}\,(\partial_\mu\Br_b)   ~~~, \cr
\rD_a\,\bU_\mu &= i(\gamma^5 \gamma_\mu)_a^{~b}\,\BB_b  - {i}(
\gamma^5\gamma_\mu \gamma^\nu)_a^{~b}\,\big(\partial_\nu \Bz_b 
\big)   \cr
&\quad- i (\gamma^5)_a^{~b}\big(\partial_\mu\Br_b \big) -\frac{i}{2}(
\gamma^5 \gamma^\nu\gamma_\mu)_a^{~b}\,\big(\partial_\nu \Br_b 
\big)  ~~~, \cr
\rD_a\,\bV_\mu &= - (\gamma_\mu)_a^{~b}\,\BB_b - (\gamma_\mu 
\gamma^\nu )_a^{~b}\,\big(\partial_\nu \Bz {}_b \big)   ~~~, \cr
&{~\,~~} + \big(\partial_\mu \Br_a \big) +\frac{1}{2}(\gamma^\nu
\gamma_\mu)_a^{~b}\,\big(\partial_\nu \Br_b \big)   ~~~, \cr          
}   \ee
for the bosons and
\be
\eqalign{
\rD_a\,\Bz_b &= -i(\gamma^\mu)_{ab}\,(\partial_\mu\bK) - (\gamma^5 
\gamma^\mu)_{ab}\,(\partial_\mu\bL) \cr
&\quad -\frac{1}{2}( \gamma^5 \gamma^\mu)_{ab}\,\bU_\mu + 
\frac{i}{2}(\gamma^\mu)_{ab}\,\bV_\mu    ~~~, \cr
\rD_a\,\Br_b &= i C_{ab}\,\bM  + (\gamma^5)_{ab}\,\bN +\frac{1}{2}(
\gamma^5 \gamma^\mu)_{ab}\,\bU_\mu +\frac{i}{2}(\gamma^\mu)_{
ab}\,\bV_\mu ~~~, \cr
\rD_a\,\BB_b &= \frac{i}{2} (\gamma^\mu)_{ab}\,(\partial_\mu\bM)
+ \frac{1}{2} (\gamma^5\gamma^\mu)_{ab}\,(\partial_\mu\bN) \cr
&\quad+ \frac{1}{2} (\gamma^5\gamma^\mu\gamma^\nu)_{ab}\,(
\partial_\mu\bU_\nu) + \frac{1}{4} (\gamma^5\gamma^\nu\gamma^\mu
)_{ab}\,(\partial_\mu\bU_\nu) \cr
&\quad+ \frac{i}{2} (\gamma^\mu\gamma^\nu)_{ab}\,(\partial_\mu\bV_\nu)
+ \frac{i}{4} (\gamma^\nu\gamma^\mu)_{ab}\,(\partial_\mu\bV_\nu) \cr
&\quad+\eta^{\mu\nu}\partial_\mu\partial_\nu(-iC_{ab}\,\bK +(\gamma^5
)_{ab}\,\bL)  ~~~, \cr          
}   \ee
for the fermions.  The action in (\ref{CLSact}) makes it clear that the Majorana
fermions $\Br_b$ and $\BB_a$ along with the scalar $\bM$, pseudoscalar
$\bN$, vector $\,\bV_\mu$ and axialvector $\,\bU_\mu$ bosonic fields
all possess equations of motion that set them identically and immediately
(i.\ e.\ without requiring further algebraic manipulations) to zero.  These
are thus the auxiliary fields of the complex linear supermultiplet.

\newpage
\section{Commutator Relations \& The Fermionic Holoraumy Tensor}
Calculations carried out algebraically and verified by Mathematica code lead to
the following results for the commutators evaluated on each field in the
CLS.  For the fermionic fields we obtain
\be \eqalign{
[\, \rD_a ~,~ \rD_b \,]\,\Bz_c &= -i C_{ab}\,(2\BB_c + (\gamma^\mu)_c^{
 ~d}\partial_\mu\Br_d) \cr
&\quad  + i(\gamma^5)_{ab}\,\big(2(\gamma^5)_c^{~d}\BB_d + (
\gamma^5\gamma^\mu)_c^{~d}\partial_\mu\Br_d\big)  \cr
&\quad  + i \, 2 (\gamma^5\gamma^\mu)_{ab}\,(\gamma^5)_c^{~d}
\partial_\mu\Bz_d   ~~~, \cr
[\, \rD_a ~,~ \rD_b \,]\,\Br_c &=-2i C_{ab}\,(\gamma^\mu)_c^{~d}
\partial_\mu\Bz_d + i \, 2 (\gamma^5)_{ab}\,(\gamma^5\gamma^\mu)_c^{
~d}\partial_\mu\Bz_d  \cr
&\quad  + i(\gamma^5\gamma^\nu)_{ab}\,\big(2(\gamma^5\gamma_\nu
)_c^{~d}\BB_d + (\gamma^5\gamma^\mu\gamma_\nu)_c^{~d}\partial_\mu
\Br_d\big)   ~~~, \cr
[\rD_a,\rD_b]\,\BB_c   &= -i C_{ab}\,\eta^{\mu\nu}\partial_\mu\partial_\nu
\Bz_c - i(\gamma^5)_{ab}\,(\gamma^5)_c^{~d}\eta^{\mu\nu}\partial_\mu
\partial_\nu\Bz_d \cr
&\quad  - i(\gamma^5\gamma^\nu)_{ab}\,(\gamma^5\gamma_\nu
\gamma^\mu)_c^{~d}\partial_\mu\BB_d  \cr
&\quad   - i(\gamma^5\gamma^\nu)_{ab}\,\big( (\gamma^5
\gamma^\mu)_c^{~d}\partial_\mu\partial_\nu \Br_d - \frac{1}{2}(
\gamma^5\gamma_\nu)_c^{~d}  \eta^{\mu\sigma} \partial_\mu
\partial_\sigma\Br_d\big) ~~~.
}  \label{HolF}
\ee
It is apparent from these equations that we may introduce an ``iso-spin''
index $\Hat I$ (where $\Hat I$ = $\Hat 1$, $\Hat 2$, or $\Hat 3$) such that
\be
{\bm \Psi}{}_a{}^{\Hat 1} ~\equiv~ \Bz_a ~~~,~~~ 
{\bm \Psi}{}_a{}^{\Hat 2} ~\equiv~ \Br_a ~~~,~~~ 
{\bm \Psi}{}_a{}^{\Hat 3} ~\equiv~ \BB_a ~~~,
\ee
and collectively write the results in (\ref{HolF}) in the form
\be \eqalign{
[\, \rD_a ~,~ \rD_b \,]\, {\bm \Psi}{}_c{}^{\Hat I} ~=~  &\left[{ {\bm H}}
{}^{(CLS)}{}^{\Hat I}{}_{\Hat K}\right]{}_{a \, b \, c}{}^d \, {\bm \Psi}{}_d{}^{
\Hat K} ~+~ \left[{ {\bm H}}{}^{\mu}{}^{(CLS)}{}^{\Hat I}{}_{\Hat K} \right]
{}_{a \, b \, c}{}^d \, \pa_{\mu} \, {\bm \Psi}{}_d{}^{\Hat K} ~   \cr
&+~  
\left[{ {\bm H}}{}^{\mu \, \nu}{}^{(CLS)} {}^{\Hat I}{}_{\Hat K} \right]{}_{a \, b \, c}{}^d \, 
\pa_{\mu} \,   \pa_{\nu} \, {\bm  \Psi}{}_d{}^{\Hat K}   ~~~,
}  \label{HolG1}
\ee
where $[{ {\bm H}}{}^{(CLS)}{}^{\Hat I}{}_{\Hat K}]{}_{a \, b \, c}{}^d$, $[{{\bm 
H}}{}^{\mu}{}^{(CLS)}{}^{\Hat I}{}_{\Hat K} ]{}_{a \, b \, c}{}^d $, and $ [{ {\bm 
H}}{}^{\mu \, \nu}{}^{(CLS)} {}^{\Hat I}{}_{\Hat K} ]{}_{a \, b \, c}{}^d$ are
tensors whose entries are {\em {only}} constants.  This is not unexpected 
as the spinors that appear in equations (\ref{HCSf}), (\ref{HTSf}), and 
(\ref{HVSf}) all possess the same engineering dimensions.  Thus, it is 
simply dimensional analysis that demands the additional presence of the 
first and third terms in (\ref{HolG1}) in comparison to the forms seen in 
(\ref{HCSf}), (\ref{HTSf}), and (\ref{HVSf}).  As seen from the action in 
(\ref{CLSact}), the engineering dimensions of $\Bz_a$ and $\Br_a$ 
differ from that of $\BB_a$ by one mass unit.  

The only non-vanishing components of these tensors are easily ``read-off''
by comparing the equation in (\ref{HolG1}) with the ones in (\ref{HolF}).  This 
leads to the results shown in (\ref{HolFG}).
\be
\eqalign{
[{{\bm H}}{}^{\mu}{}^{(CLS)}{}^{\Hat 1}{}_{\Hat 1} ]{}_{a \, b \, c}{}^d 
~&=~ i \, 2 (\gamma^5\gamma^\mu)_{ab}\,(\gamma^5)_c^{~d}
~~~, \cr
[{{\bm H}}{}^{\mu}{}^{(CLS)}{}^{\Hat 1}{}_{\Hat 2} ]{}_{a \, b \, c}{}^d 
~&=~ -i C_{ab}\, (\gamma^\mu)_c^{~d} + i(\gamma^5)_{ab} \,  (
\gamma^5\gamma^\mu)_c^{~d}      ~~~,  \cr
[{ {\bm H}}{}^{(CLS)}{}^{\Hat 1}{}_{\Hat 3}]{}_{a \, b \, c}{}^d  ~&=~
- i \, 2 C_{ab}\, ({\bm {\rm I}_4)}{}_c^{ ~d} + i \, 2(\gamma^5)_{ab} \,(\gamma^5
)_c^{~d}  ~~~,  \cr
[{{\bm H}}{}^{\mu}{}^{(CLS)}{}^{\Hat 2}{}_{\Hat 1} ]{}_{a \, b \, c}{}^d 
~&=~ -2i C_{ab}\,(\gamma^\mu)_c^{~d} + i \, 2 (\gamma^5)_{ab}\,(
\gamma^5\gamma^\mu)_c^{~d} ~~~, \cr
[{{\bm H}}{}^{\mu}{}^{(CLS)}{}^{\Hat 2}{}_{\Hat 2} ]{}_{a \, b \, c}{}^d 
~&=~   i \, 2 (\gamma^5)_{ab}\,(\gamma^5\gamma^\mu)_c^{
~d}    ~~~,  \cr
[{ {\bm H}}{}^{(CLS)}{}^{\Hat 2}{}_{\Hat 3}]{}_{a \, b \, c}{}^d  ~&=~ i \, 2 
(\gamma^5\gamma^\nu)_{ab}\,(\gamma^5\gamma_\nu)_c^{~d}  
~~~,   \cr
[{ {\bm H}}{}^{(CLS)}{}^{\mu \, \nu}{}^{\Hat 3}{}_{\Hat 1}]{}_{a \, b \, c}{}^d  
~&=~ -i C_{ab}\,\eta^{\mu\nu} ({\bm {\rm I}}{}_4)_c^{~d} - i(\gamma^5)_{ab}\,(
\gamma^5)_c^{~d} \eta^{\mu\nu} ~~,  \cr
[{ {\bm H}}{}^{\mu \, \nu}{}^{(CLS)} {}^{\Hat 3}{}_{\Hat 2} ]{}_{a \, b \, c}{}^d 
~&=~  - i(\gamma^5\gamma^\nu)_{ab}\, (\gamma^5 \gamma^\mu)_c^{
~d} + i \,  \frac{1}{2} (\gamma^5\gamma^\rho)_{ab}\, (\gamma^5
\gamma_\rho)_c^{~d}  \eta^{\mu\nu} ~~,  \cr
[{ {\bm H}}{}^{\mu}{}^{(CLS)} {}^{\Hat 3}{}_{\Hat 3} ]{}_{a \, b \, c}{}^d 
~&=~ - i(\gamma^5\gamma^\nu)_{ab}\,(\gamma^5\gamma_\nu
\gamma^\mu)_c^{~d}  ~~,
}  
\label{HolFG}
\ee
where all other components of these tensor equal to zero.  These
expressions extend the concept of the four dimensional fermionic
holoraumy \cite{HoLoRmY4D} tensor to cover the case of the complex 
linear supermultiplet.

For the bosonic fields we obtain, 
\be \eqalign{
[\, \rD_a ~,~ \rD_b \,]\,\bK &= 2i C_{ab}\,\bM  + 2(\gamma^5)_{ab}\,\bN
+ 2(\gamma^5\gamma^\nu)_{ab}\,(\bU_\nu + \partial_\nu\bL) ~~~, \cr
[\, \rD_a ~,~ \rD_b \,]\,\bL &= 2i C_{ab}\,\bN - 2(\gamma^5)_{ab}\,\bM
 + 2(\gamma^5\gamma^\nu)_{ab}\,(\bV_\nu - \partial_\nu\bK) ~~~, \cr
[\, \rD_a ~,~ \rD_b \,]\,\bM &= 2i C_{ab}\,\eta^{\mu\nu}\partial_\mu(\bV_\nu 
- \partial_\nu\bK) \cr
&\quad  + 2(\gamma^5)_{ab}\,\eta^{\mu\nu}\partial_\mu(\bU_\nu + 
\partial_\nu\bL) + 2(\gamma^5\gamma^\nu)_{ab}\,\partial_\nu\bN  
~~~, \cr
[\, \rD_a ~,~ \rD_b \,]\,\bN &=-2i C_{ab}\,\eta^{\mu\nu}\partial_\mu(\bU_\nu 
+ \partial_\nu\bL) \cr
&\quad  + 2(\gamma^5)_{ab}\,\eta^{\mu\nu}\partial_\mu(\bV_\nu - 
\partial_\nu\bK) - 2(\gamma^5\gamma^\nu)_{ab}\,\partial_\nu\bM  
~~~, \cr
[\rD_a,\rD_b]\,\bU_\mu &=-4i C_{ab}\,\partial_\mu\bN + 4(\gamma^5
)_{ab}\,\partial_\mu\bM - 2(\gamma^5\gamma^\nu)_{ab}\,\big( 
\partial_\mu \bV_\nu  \big)  \cr
&\quad  + 2 (\gamma^5\gamma^\nu)_{ab}\,\big( \eta_{\mu\nu}\eta^{
\rho\sigma}\partial_\rho(2\partial_\sigma \bK - \bV_\sigma) + 
\epsilon^{\rho\sigma}_{~~\mu\nu}\partial_\rho \bU_\sigma\big)  
~~~, \cr
[\rD_a,\rD_b]\,\bV_\mu &= 4i C_{ab}\,\partial_\mu\bM + 4(\gamma^5
)_{ab}\,\partial_\mu\bN + 2(\gamma^5\gamma^\nu)_{ab}\big(
\partial_\mu \bU_\nu   \big)   \cr
&\quad  + 2 (\gamma^5\gamma^\nu)_{ab}\,\big(  \eta_{\mu\nu}\eta^{
\rho\sigma}\partial_\rho(2\partial_\sigma \bL + \bU_\sigma) -  \epsilon^{
\rho\sigma}_{~~\mu\nu}\partial_\rho \bV_\sigma\big) ~~~.
}  \label{HolB}
\ee

For the bosonic fields in (\ref{HolB}), it is possible to also define a 
new single bosonic field variable analogous ${\bm \Psi}{}_c{}^{\Hat 
I}$ to that carries an ``iso-spin'' index.  However, we have found no
evidence from our past investigations of the concept of the bosonic
holoraumy that using such an extended structure is useful.  Therefore, 
we forego doing so.
  
The results in this chapter depend on substantial use of Fierz identities.
In order to ensure best practices and the correctness of these intricate
calculations, several teams were created to calculate each of the 
expressions for the commutators.  Each team worked {\em {independently}}
of one another and only at the completion of the calculation of a
commutator were the results checked.  Any discrepancies were then
used to ``de-bug'' the algebraic calculations leading to divergent
results.  The final results presented in this chapter are the consensus
ones arising from this process.

\newpage
\section{The 0-Brane Formulation}
\label{s4b}

To obtain the ``0-brane" formulation of the CLS, we restrict the coordinate 
dependence of all the fields so that they only depend on the temporal 
coordinate.  Accordingly, this leads to the following form of the super
supersymmetry transformation laws 
\be
\eqalign{
\rD_a\,\Bz_b &= -i(\gamma^0)_{ab}\,(\partial_{\t} \, \bK) - (\gamma^5 
\gamma^0)_{ab}\,(\partial_{\t}\bL) \cr
&\quad -\frac{1}{2}( \gamma^5 \gamma^\mu)_{ab}\,\bU_\mu + \frac{i}{2}
(\gamma^\mu)_{ab}\,\bV_\mu    ~~~, \cr
\rD_a\,\Br_b &= i C_{ab}\,\bM  + (\gamma^5)_{ab}\,\bN +\frac{1}{2}( 
\gamma^5 \gamma^\mu)_{ab}\,\bU_\mu +\frac{i}{2}(\gamma^\mu)_{
ab}\,\bV_\mu ~~~, \cr
\rD_a\,\BB_b &= \frac{i}{2} (\gamma^0)_{ab}\,(\pa_{\t}\bM) + \frac{1}{2} 
(\gamma^5\gamma^0)_{ab}\,(\pa_{\t}\bN) \cr
&\quad+ \frac{1}{4} (\gamma^5\gamma^0\gamma^\nu)_{ab}\,(\pa_{\t}\bU_\nu)
  + \frac{i}{4} (\gamma^0\gamma^\nu)_{ab}\,(\pa_{\t}\bV_\nu)  \cr
&\quad +\pa_{\t}  \pa_{\t}   (iC_{ab}\,\bK - (\gamma^5)_{ab}\,\bL)  ~~~, \cr          
}   \ee
for the fermions and
\be
\eqalign{
\rD_a\,\bK &= \Br_a - \Bz_a ~~~, \cr
\rD_a\,\bL &= i(\gamma^5)_a^{~b} (\Br_b + \Bz_b) ~~~, \cr
\rD_a\,\bM &= \BB_a -\frac{1}{2} (\gamma^0)_a^{~b}\,(\pa_{\t} \Br_b)    
~~~, \cr
\rD_a\,\bN &=  -i (\gamma^5)_a^{~b}\,\BB_b +\frac{i}{2} (\gamma^5 
\gamma^0)_a^{~b}\,(\pa_{\t}\Br_b)   ~~~, \cr
\rD_a\,\bU_0 &= i(\gamma^5 \gamma_0)_a^{~b}\,\BB_b - {i}(
\gamma^5)_a^{~b}\,\big(\partial_{\t} \Bz_b \big) - i \frac{3}{2}
(\gamma^5 )_a^{~b}\,\big(\partial_{\t} \Br_b \big)  ~~~, \cr
\rD_a\,\bU_i &= i(\gamma^5 \gamma_i)_a^{~b}\,\BB_b + {i}(
\gamma^5  \gamma^0 \gamma_i)_a^{~b}\,\big(\partial_{\t} \Bz_b 
\big) -\frac{i}{2}(\gamma^5 \gamma^0 \gamma_i)_a^{~b}
\,\big(\partial_{\t} \Br_b \big)  ~~~, \cr
\rD_a\,\bV_0 &= - (\gamma_0)_a^{~b}\,\BB_b - \,\big( \pa_\t \Bz 
{}_a \big)    + \frac{3}{2}\,\big( \pa_\t \Br_a \big) ~~~, \cr
\rD_a\,\bV_i &= - (\gamma_i)_a^{~b}\,\BB_b + ( \gamma^0 
\gamma_i)_a^{~b}\,\big( \pa_\t \Bz {}_b \big)  + \frac{1}{2}(
\gamma^0 \gamma_i)_a^{~b}\,\big( \pa_\t \Br_b \big)  ~~~. \cr          
}   \ee
for the bosons.
The same restriction of the coordinate dependence for the fields
is next applied to the calculations of the commutators of the superspace
D-operators for all the fields.  We obtain the following equations
\be \eqalign{
 [\, \rD_a ~,~ \rD_b \,]\,\Bz_c &= -i C_{ab}\,(2\BB_c + (\gamma^0)_c^{
 ~d}\pa_{\t}\Br_d) \cr
&\quad  + i(\gamma^5)_{ab}\,\big(2(\gamma^5)_c^{~d}\BB_d + (
\gamma^5\gamma^0)_c^{~d}\pa_{\t}\Br_d\big)  \cr
&\quad  + i\, 2 (\gamma^5\gamma^0)_{ab}\,(\gamma^5)_c^{~d}
 \pa_\t  \Bz_d   ~~~, \cr
[\, \rD_a ~,~ \rD_b \,]\,\Br_c &=-i 2 C_{ab}\,(\gamma^0)_c^{~d}\pa_{\t}
\Bz_d + i\, 2 (\gamma^5)_{ab}\,(\gamma^5\gamma^0)_c^{~d}\pa_{\t}\Bz_d  
\cr
&\quad  + i(\gamma^5\gamma^\nu)_{ab}\,\big(2(\gamma^5\gamma_\nu
)_c^{~d}\BB_d + (\gamma^5\gamma^0\gamma_\nu)_c^{~d}\pa_{\t}\Br_d
\big)   ~~~, \cr
[\rD_a,\rD_b]\,\BB_c   &= i C_{ab}\, \pa_{\t}  \pa_{\t}   \Bz_c + i(\gamma^5
)_{ab}\,(\gamma^5)_c^{~d} \pa_{\t}\partial_\t \Bz_d \cr
&\quad  - i(\gamma^5\gamma^\nu)_{ab}\, (\gamma^5\gamma_\nu
\gamma^0)_c^{~d}\pa_{\t}\BB_d  \cr
&\quad  - i  (\gamma^5\gamma^0)_{ab}\,\big( (\gamma^5
\gamma^0 )_c^{~d} \pa_{\t}  \pa_{\t}  \Br_d     \big)      \cr
&\quad  + i  \frac{1}{2}  (\gamma^5\gamma^\nu)_{ab}\,\big( 
(\gamma^5\gamma_\nu)_c^{~d} \pa_{\t}  \pa_{\t}  \Br_d\big) ~~~.
}  \label{HolF2}
\ee
for the fermions and similar for the bosonic fields we obtain the 
equations seen below.
\be \eqalign{
[\, \rD_a ~,~ \rD_b \,]\,\bK &= 2i C_{ab}\,\bM  + 2(\gamma^5)_{ab}\,\bN
+ 2(\gamma^5\gamma^\nu)_{ab}\,(\bU_\nu) 
+ 2(\gamma^5\gamma^0)_{ab}\,( \pa_\t \bL) ~~~, \cr
[\, \rD_a ~,~ \rD_b \,]\,\bL &= 2i C_{ab}\,\bN - 2(\gamma^5)_{ab}\,\bM
+ 2(\gamma^5\gamma^\nu)_{ab}\,(\bV_\nu)
- 2(\gamma^5\gamma^0)_{ab}\,( \pa_\t \bK) ~~~, \cr
[\, \rD_a ~,~ \rD_b \,]\,\bM &= - i 2 C_{ab}\, \pa_{\t}(\bV_0 -  \pa_\t 
\bK)  - 2(\gamma^5)_{ab}\,  \pa_{\t}(\bU_0 +  \pa_\t \bL) \cr
&\quad   + 2(\gamma^5\gamma^0)_{ab}\, \pa_\t  \bN  ~~~, \cr
[\, \rD_a ~,~ \rD_b \,]\,\bN &=i 2 C_{ab}\,\pa_{\t}(\bU_0 +  \pa_\t 
\bL)  - 2(\gamma^5)_{ab}\, \pa_{\t}(\bV_0 -  \pa_\t \bK)   \cr
&\quad - 2(\gamma^5\gamma^0)_{ab}\, \pa_\t  \bM  ~~~, \cr
[\rD_a,\rD_b]\,\bU_0 &=-4i C_{ab}\,\pa_{\t}\bN + 4(\gamma^5)_{ab}
\,\pa_{\t}\bM -2(\gamma^5\gamma^i)_{ab}\,\big(\pa_{\t}\bV_i
\big) \cr
&\quad  +  4(\gamma^5\gamma^0)_{ab}   \, \big(  \pa_\t (\pa_\t
\bK - \bV_0 )  \big)  ~~~, \cr  
[\rD_a,\rD_b]\,\bU_i &=  - 2(\gamma^5\gamma_i)_{ab}\,\big(  \pa_\t 
(2 \pa_\t  \bK - \bV_0) \big) + 2(\gamma^5 \gamma^j)_{ab}\,\big(  
\epsilon_{i j}{}^k  \pa_\t \bU_k \big)  ~~~, \cr
[\rD_a,\rD_b]\,\bV_0 &= 4i C_{ab}\,\pa_{\t} \bM + 4(\gamma^5)_{ab}
\,\pa_{\t}\bN + 2(\gamma^5\gamma^i)_{ab}\big(\pa_{\t}\bU_i
\big)   \cr
&\quad  +  4(\gamma^5\gamma^0)_{ab} \big( \pa_\t ( \pa_\t
\bL + \bU_0)     \big) ~~~, \cr
[\rD_a,\rD_b]\,\bV_i &=  - 2(\gamma^5\gamma_i )_{ab}  \big(
 \pa_\t (2 \pa_\t \bL + \bU_0) \big)  -  2(\gamma^5
\gamma^j)_{ab}  \big( \epsilon_{i \, j}{}^k  \pa_\t \bV_k \big) ~~~.
}  \label{HolB2}
\ee
Since these results are expressed in terms of the gamma matrices
for the four dimensional theory. information about the Lorentz
symmetry properties of the various fields are still apparent.

\newpage
\section{Obtaining The Valise Formulation}
\label{s3b}

The reduction process in the last section can be used as the starting point to
derive a valise 0-brane formulation via some ``field redefinitions.''
These redefinitions take the forms,
\be
\bM ~\to ~ \pa_{\t}  \bM  ~~,~~ \bN ~\to ~ \pa_{\t}  \bN ~~,~~ \bU_\mu 
 ~\to ~ \pa_{\t} \bU_\mu  ~~,~~  \bV_\mu ~\to ~ \pa_{\t} \bV_\mu
 ~~,~~  {\bs\beta}_a  ~\to ~   \pa_{\t}  {\bs\beta}_a  ~~.
\ee
A characteristic of a valise formulation is that when expressed after the
field redefinitions all the bosons possess the same engineering dimension
and all the fermions possess the same engineering dimension, though
the latter differs from that of the bosons.  This is reflected in the fact that
the valise formulation of the action takes the form
\be
\eqalign{  
\mathcal{L}_{\text{\CLS}}^{V} &= \frc{1}2  \left(  \pa_\t \bs K \right){}^2  
~+~ \frc{1}2  \left(  \pa_\t \bs L \right){}^2  ~-~ \frc{1}2  \left(  \pa_\t 
\bs M \right){}^2 ~-~  \frc{1}2  \left(  \pa_\t \bs N \right){}^2  ~    \cr
&~~~\, +~ \frc{1}4\,  \left(  \pa_\t  {\bs U}_i \right)\, \left(  \pa_\t  {\bs U
}_i \right)   ~-~ \frc{1}4\,  \left(  \pa_\t  {\bs U}_0 \right)\, \left(  \pa_\t  {\bs U
}_0 \right)    \cr
&~~~\, +~  \frc{1}4\,  \left(  \pa_\t  {\bs V}_i \right)\, \left(  \pa_\t 
{\bs V}_i \right)  ~-~  \frc{1}4\,  \left(  \pa_\t  {\bs V}_0 \right)\, \left(  \pa_\t 
{\bs V}_0 \right)   \cr
& ~~~\,+~  i \frc{1}2 (\gamma^0)^{ab} \, {\bs \zeta}_a \,  \pa_\t {\bs \zeta}_b  
~+~ i\,  C^{ab} \, {\bs \rho}_a \,   \pa_\t  \, {\bs\beta}_b ~~~,
}  \label{CLSact1} \ee
which is accommodated by being invariant (once more up to a total
derivative) under a set of modified supersymmetry transformation
laws.  For the field variables that appear in (\ref{CLSact1}) these take
 the forms
\be
\eqalign{
\rD_a\,\bK &= \Br_a - \Bz_a ~~~, \cr
\rD_a\,\bL &= i(\gamma^5)_a^{~b} (\Br_b + \Bz_b) ~~~, \cr
\rD_a\,\bM &= \BB_a -\frac{1}{2} (\gamma^0)_a^{~b}\,  \Br_b    
~~~, \cr
\rD_a\,\bN &=  -i (\gamma^5)_a^{~b}\,\BB_b +\frac{i}{2} (\gamma^5 
\gamma^0)_a^{~b}\, \Br_b   ~~~, \cr
\rD_a\,\bU_0 &= i(\gamma^5 \gamma_0)_a^{~b}\,\BB_b - {i}(
\gamma^5)_a^{~b}\,  \Bz_b  - i \frac{3}{2}
(\gamma^5 )_a^{~b}\,  \Br_b   ~~~, \cr
\rD_a\,\bU_i &= i(\gamma^5 \gamma_i)_a^{~b}\,\BB_b + {i}(
\gamma^5  \gamma^0 \gamma_i)_a^{~b}\,  \Bz_b 
 -\frac{i}{2}(\gamma^5 \gamma^0 \gamma_i)_a^{~b}
\,  \Br_b   ~~~, \cr
\rD_a\,\bV_0 &= - (\gamma_0)_a^{~b}\,\BB_b -  \Bz 
{}_a     + \frac{3}{2}\,  \Br_a  ~~~, \cr
\rD_a\,\bV_i &= - (\gamma_i)_a^{~b}\,\BB_b + ( \gamma^0 
\gamma_i)_a^{~b}\,  \Bz {}_b   + \frac{1}{2}(
\gamma^0 \gamma_i)_a^{~b}\, \Br_b  ~~~, \cr
\rD_a\,\Bz_b &= -i(\gamma^0)_{ab}\,(\partial_{\t} \, \bK) - (\gamma^5 
\gamma^0)_{ab}\,(\partial_{\t}\bL) \cr
&\quad -\frac{1}{2}( \gamma^5 \gamma^\mu)_{ab}\,
(\partial_{\t} \, \bU_\mu ) + \frac{i}{2}
(\gamma^\mu)_{ab}\, (\partial_{\t} \,\bV_\mu )   ~~~, \cr
\rD_a\,\Br_b &= i C_{ab}\, (\partial_{\t} \, \bM ) + (\gamma^5)_{ab}\, 
(\partial_{\t} \, \bN ) + \frac{1}{2}( \gamma^5 \gamma^\mu)_{ab}\,
(\partial_{\t} \, \bU_\mu ) +\frac{i}{2}(\gamma^\mu)_{
ab}\, (\partial_{\t} \, \bV_\mu ) ~~~, \cr
\rD_a\,\BB_b &= \frac{i}{2} (\gamma^0)_{ab}\,(\pa_{\t}\bM) + \frac{1}{2} 
(\gamma^5\gamma^0)_{ab}\,(\pa_{\t}\bN) \cr
&\quad+ \frac{1}{4} (\gamma^5\gamma^0\gamma^\nu)_{ab}\,(\pa_{\t}
\bU_\nu)  + \frac{i}{4} (\gamma^0\gamma^\nu)_{ab}\,(\pa_{\t}\bV_\nu)  \cr
&\quad +   \pa_{\t}   (iC_{ab}\,\bK - (\gamma^5)_{ab}\,\bL)  ~~~. \cr          
} \label{VHVH}  \ee

These lead to the following expressions for the fermionic fields.
\be \eqalign{
 [\, \rD_a ~,~ \rD_b \,]\,\Bz_c &= -i C_{ab}\,(2 (\partial_{\t} \,
 \BB_c ) + (\gamma^0)_c^{ ~d} (\pa_{\t}\Br_d) ~) \cr
&\quad  + i(\gamma^5)_{ab}\,\big(2(\gamma^5)_c^{~d} (\partial_{\t} \,
\BB_d )+ (
\gamma^5\gamma^0)_c^{~d} \, (\pa_{\t}\Br_d )  \cr
&\quad  + i\, 2 (\gamma^5\gamma^0)_{ab}\,(\gamma^5)_c^{~d} \, 
( \pa_\t  \Bz_d )  ~~~, \cr
[\, \rD_a ~,~ \rD_b \,]\,\Br_c &=-i 2 C_{ab}\,(\gamma^0)_c^{~d} \,
(\pa_{\t} \Bz_d ) + i\, 2 (\gamma^5)_{ab}\,(\gamma^5\gamma^0)_c^{~d}
\, (\pa_{\t}\Bz_d  ) \cr
&\quad  + i(\gamma^5\gamma^\nu)_{ab}\,\big(2(\gamma^5\gamma_\nu
)_c^{~d}\, ( \pa_\t \BB_d )+ (\gamma^5\gamma^0\gamma_\nu)_c^{~d}
\, ( \pa_{\t}\Br_d )   ~~~, \cr
[\rD_a,\rD_b]\,\BB_c   &= i C_{ab}\, (\pa_{\t}   \Bz_c ) + i(\gamma^5
)_{ab}\,(\gamma^5)_c^{~d} \, (\partial_\t \Bz_d ) \cr
&\quad  - i(\gamma^5\gamma^\nu)_{ab}\, (\gamma^5\gamma_\nu
\gamma^0)_c^{~d} \, ( \pa_{\t}\BB_d ) \cr
&\quad  - i  (\gamma^5\gamma^0)_{ab}\, (\gamma^5
\gamma^0 )_c^{~d} \, ( \pa_{\t}  \Br_d  )      \cr
&\quad  + i  \frac{1}{2}  (\gamma^5\gamma^\nu)_{ab}\,\big( 
(\gamma^5\gamma_\nu)_c^{~d} \, ( \pa_{\t}  \Br_d ) ~~~.
}  \label{HolF2}
\ee
For the bosonic fields we obtain the following expressions below.
\be \eqalign{ {~~~~~~}
[\, \rD_a ~,~ \rD_b \,]\,\bK &= 2i C_{ab}\,( \pa_\t \bM ) + 2(\gamma^5
)_{ab}\, ( \pa_\t \bN ) + 2(\gamma^5\gamma^\nu)_{ab}\,(\pa_\t
\bU_\nu) 
+ 2(\gamma^5\gamma^0)_{ab}\,( \pa_\t \bL) ~~~, \cr
[\, \rD_a ~,~ \rD_b \,]\,\bL &= 2i C_{ab}\, (\pa_\t \bN)  - 2(\gamma^5
)_{ab}\, ( \pa_\t \bM )
+ 2(\gamma^5\gamma^\nu)_{ab}\,( \pa_\t \bV_\nu)
- 2(\gamma^5\gamma^0)_{ab}\,( \pa_\t \bK) ~~~, \cr
[\, \rD_a ~,~ \rD_b \,]\,\bM &= - i 2 C_{ab}\,( \pa_{\t}(\bV_0 -  
\bK) ) - 2(\gamma^5)_{ab}\, ( \pa_{\t}(\bU_0 +   \bL) ) \cr
&\quad   + 2(\gamma^5\gamma^0)_{ab}\, (\pa_\t  \bN)  ~~~, \cr
[\, \rD_a ~,~ \rD_b \,]\,\bN &=i 2 C_{ab}\, ( \pa_{\t}(\bU_0  +  
\bL) ) - 2(\gamma^5)_{ab}\, (\pa_{\t}(\bV_0 -  \bK) )  \cr
&\quad - 2(\gamma^5\gamma^0)_{ab}\, (\pa_\t  \bM  ) ~~~, \cr
[\rD_a,\rD_b]\,\bU_0 &=-4i C_{ab}\, (\pa_{\t}\bN ) + 4(\gamma^5)_{ab}
\, ( \pa_{\t}\bM)  - 2(\gamma^5\gamma^i)_{ab}\,\big(\pa_{\t}\bV_i
\big) \cr
&\quad  +  4(\gamma^5\gamma^0)_{ab}   \, \big(  \pa_\t (
\bK - \bV_0 )  \big)  ~~~, \cr  
[\rD_a,\rD_b]\,\bU_i &=  - 2(\gamma^5\gamma_i)_{ab}\,\big(  \pa_\t 
(2  \bK - \bV_0) \big) + 2(\gamma^5 \gamma^j)_{ab}\, \epsilon_{i j
}{}^k \, (  \pa_\t \bU_k )  ~~~, \cr
[\rD_a,\rD_b]\,\bV_0 &= 4i C_{ab}\, ( \pa_{\t} \bM ) + 4(\gamma^5)_{ab}
\, (\pa_{\t}\bN ) + 2(\gamma^5\gamma^i)_{ab}\big(\pa_{\t}\bU_i
\big)   \cr
&\quad  +  4(\gamma^5\gamma^0)_{ab} \big( \pa_\t ( 
\bL + \bU_0)     \big) ~~~, \cr
[\rD_a,\rD_b]\,\bV_i &=  - 2(\gamma^5\gamma_i )_{ab}  \big(
 \pa_\t (2  \bL + \bU_0) \big)  -  2(\gamma^5
\gamma^j)_{ab}  \, \epsilon_{i \, j}{}^k \, ( \pa_\t \bV_k \big) ~~~.
}  \label{HolB2}
\ee

These equations in (\ref{VHVH}) are similar in form to those that appear 
in (\ref{VH1}).  These equation can in fact be used to derived an adinkra 
formulation of the complex linear supermultiplet and its associated $\bm 
{\Tilde V}$-matrices.  But in fact the equations in (\ref{VH1}) and (\ref{VHVH})
are totally distinct in the information each set carries.

The equations in (\ref{VHVH}) carry information about the adjacency
matrices of the underlying adinkra network which can be the 1D, four
color ``shadow'' of a 4D, $\cal N$ = 1 supermultiplet.  However, these
equations carry no obvious information about the Lorentz symmetry
of the 4D, $\cal N$ = 1 supermultiplet they may shadow. 

On the other hand, the equations in (\ref{VHVH}) carry information
about the Lorentz symmetry of the 4D, $\cal N$ = 1 complex linear
supermultiplet, but carry no obvious information about the adjacency
matrices of the shadow.

\newpage
\section{Documentation On Codes Utilized}
\label{codesUsd}  

Here we wish to provide documentation on some of the codes used
in this work to derive and verify the results that have been presented 
in chapter five.  The actual codes themselves are available to any 
interested party by clicking the hyperlink available at
 
https://github.com/vkorotkikh/SUSY-4DHoloraumy-4DN1CL-Supermultiplet
\newline $~$ 
\newline  \noindent
on-line.  Two separate sets of codes were used in deriving the results.
  
\subsection{Description of Code to Verify Commutators}

A Mathematica notebook Coefficient Check.nb was constructed to determine 
the right hand side coefficient, designated below as C${}_0$, such that the 
following equation holds true:
\begin{center}
$\frac{i}{2}\eta _{\nu \sigma }\left ( \gamma ^{5}\gamma ^{\sigma }\gamma ^{5}\gamma ^{\mu } \right )^{c}{_{e}}\left ( \gamma ^{5} \right ){_{c}}^{f}\partial_{\mu}\zeta _{f}-\frac{i}{2}\eta _{\nu \sigma }\left ( \gamma ^{5}\gamma ^{\sigma }\gamma ^{5}\gamma ^{\mu } \right )^{c}{_{e}}\left ( \gamma ^{5} \right ){_{c}}^{f}\partial_{\mu}\zeta _{f}+\frac{i}{4}\eta _{\nu \sigma }\left ( \gamma ^{5}\gamma ^{\sigma }\gamma ^{5}\gamma ^{\mu } \right )^{c}{_{e}}\left ( \gamma ^{5}\gamma ^{\rho}\gamma _{\mu} \right ){_{c}}^{f}\partial_{\rho}\zeta _{f}-\frac{i}{4}\eta _{\nu \sigma }\left ( \gamma ^{5}\gamma ^{\sigma }\gamma ^{5}\gamma ^{\mu } \right )^{c}{_{e}}\left ( \gamma ^{\rho}\gamma _{\mu} \right ){_{c}}^{f}\partial_{\rho}\zeta _{f}={\rm C}{}_0\left ( \gamma ^{5} \right ){_{e}}^{f}\partial_{\nu}\zeta _{f}$
\end{center}
This was a necessary step in confirming the coefficients of the Complex Linear Supermultiplet zeta commutator relation. The code confirmed that $2i$ was the correct coefficient for the right hand side of the equation above.
\\
\\
\indent
The code begins by clearing all associations using the $ClearAll$["\`{}*"] command and initializing the notation pallet. The Notation Palette allows users to associate an external representation, for example a set of symbols or characters, to a particular internal representation of symbols, variables, or functions. The notation pallet is used here to construct indexable symbolic super derivative operators, and indexable symbolic zetas. This is done by creating functions of variables that can be looped over, and assigning those variables as corresponding subscripts to another set of symbols. Additionally other useful objects are also defined, this includes the $2x2$ identity matrix, the gamma matrices, and the eta and spinor metric.
\\
\\
\indent
Next, using Mathematica's Part function, $\left [ \left [ \_\right ] \right ]$ to account for indices, the left hand side of the above equation is constructed. In Mathematica the left hand side of this equation is described as a function of many variables, each of which correspond to a tensor index. These variables are selected via the part function, and can be looped over to draw from the definitions in the first section of the code.
\\
\\
\indent
A $Table$ loop is then used to symbolically determine all the values for the left hand side of the equation. The result is a $\{4 \times 4 \times 4 \times 4 \times 4 \times 4 \times 4\}$ matrix of super derivative operators and zetas with coefficients 1/4 and -1/4, as well as many 0 elements. Then using another $Table$ loop and the $Sum$ function, the previously determined values are summed over according to the summations originally prescribed by the indicies in the left hand side of the above equation. The result is a $\{4 \times 4\}$ solution matrix for the left hand side of the above equation containing superderivative operators and zetas, all of which have coefficient $-2$ or $2$.
\\
\\
\indent
The same process is then conducted for the right hand side of the equation. A right hand side solution matrix emerges with all the resulting combinations of superderiviative operators and zetas. However this time, all the coefficients are $-i$, or $i$. 
\\
\\
\indent
Now if we compare each $\partial{_{\mu }}^{th} \, \zeta {_{f}}^{th}$ element of the left hand side solution matrix and the right hand side solution matrix, disregarding their position in either matrix, as this varies due to how the indicies are summed over in Mathematica. We see that if each $\partial{_{\mu }}^{th} \, \zeta {_{f}}^{th}$ element in the right hand side solutions matrix, if multiplied by $2i$, it is equal to it's corresponding $\partial{_{\mu }}^{th} \, \zeta {_{f}}^{th}$ element in the left hand side solution matrix.
\\
\\
\indent
This shows that $LHS = i \, 2\,RHS$. Thus the right hand side C${}_0$ coefficient is $2i$. A quick explicit check using the $Equal$ function is performed solidifying the result.

This code may be adapted for other similar calculations by any user
familiar with Mathematica.

\subsection{Description of Coefficient Check.nb}

If the goal is to show that our expressions for the commutators on the Complex Linear Supermultiplet are correct, then it is sufficient to show that they both expand to the same linear combination of space-time derivatives of component fields since the set of derivatives is linearly independent. Thus, the code must efficiently and correctly expand commutator expressions and sums over both spinor indices as well as space-time indices. Mathematica is perfect for these sorts of calculations due to its incredibly powerful symbolic manipulation faculties.

The code begins with a number of notation definitions; this is not strictly necessary, but makes the code much more readable. Then follows a number of definitions of our conventions for spinor and space-time metrics as well as the $\gamma$-matrices. Once these definitions are established, we can then proceed to giving Mathematica replacement rules to use to simply commutator expressions. In order to prevent conflicts with built-in Mathematica operators, we use $
\vCent
{\setlength{\unitlength}{1mm}
\begin{picture}(-20,0)
\put(-2.2,-1.66){\includegraphics[width=0.136in]{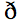}}
\end{picture}}
$
$\,$ and $
\vCent
{\setlength{\unitlength}{1mm}
\begin{picture}(-20,0)
\put(-2.2,-1.44){\includegraphics[width=0.16in]{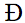}}
\end{picture}}
$
$\,\,$ as the symbolic derivatives. For each index of $
\vCent
{\setlength{\unitlength}{1mm}
\begin{picture}(-20,0)
\put(-2.2,-1.66){\includegraphics[width=0.136in]{DHat}}
\end{picture}}
$
$\,$ and \newline \noindent $
\vCent
{\setlength{\unitlength}{1mm}
\begin{picture}(-20,0)
\put(-2.2,-1.36){\includegraphics[width=0.16in]{DBar}}
\end{picture}}
$
$~\,$ 
we define them to be linear and to commute with each other. It is important to note that these definitions are one-way, i.e. $
\vCent
{\setlength{\unitlength}{1mm}
\begin{picture}(-20,0)
\put(-2.2,-2.0){\includegraphics[width=1.7in]{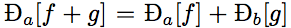}}
\end{picture}}
$
$~\,$ 
$~~~~~~~~~~~~~~~~~~~~~~~~~~~$
but not the other way around. This is to prevent the \textit{Simplify} function from entering an infinite loop.

Once the basic properties of $
\vCent
{\setlength{\unitlength}{1mm}
\begin{picture}(-20,0)
\put(-2.26,-1.66){\includegraphics[width=0.136in]{DHat}}
\end{picture}}
$
$\,$ and $
\vCent
{\setlength{\unitlength}{1mm}
\begin{picture}(-20,-1)
\put(-2.46,-1.36){\includegraphics[width=0.16in]{DBar}}
\end{picture}}
$
$\,$ are defined, the replacement rules for $
\vCent
{\setlength{\unitlength}{1mm}
\begin{picture}(-20,0)
\put(-2.2,-1.66){\includegraphics[width=0.136in]{DHat}}
\end{picture}}
$
$\,$ on each component field are specified. These rules are simply the definitions of the Complex Linear Supermultiplet. With these in hand, we can now proceed to the actual verification of the calculations. The \textit{Simplify} command can now take our final expressions for each commutator and use the replacement rules specified above to break each expression down into a linear combination of space-time derivatives of component fields and verify that both sides are equivalent (or not).

This technique isn't just limited to the Complex Linear Supermultiplet, nor calculating commutators. Variants of this program could evaluate any expression, commutators or otherwise, of any supermultiplet simply by replacing the definitions of the differential operator $
\vCent
{\setlength{\unitlength}{1mm}
\begin{picture}(-20,0)
\put(-2.2,-1.66){\includegraphics[width=0.136in]{DHat}}
\end{picture}}
$
$\,$ with its action on any other supermultiplet.

\newpage
\section{Conclusions}
\label{conclusions}

 \vskip,2in
 
To state most clearly the implications of our results, we first note the data 
used to find the weights and roots of ordinary Lie algebra representations
are accessed through the eigenvalues of the matrices that represent the
maximal commuting set of generators.  This maximal commuting set arises 
due to the applicability of the Jordan-Chevalley decomposition to Lie algebras.  
The Jordan-Chevalley decomposition, as suggested by H\" ubsch, seems 
{\em {not}} to apply to the algebra of 1D, $N$-extended supersymmetry.  
Accordingly, to obtain the equivalent data from representations of the 1D, 
$N$-extended supersymmetry  algebra, one is forced to look elsewhere.  It 
is also our contention that the 4D, $\cal N$ = 1 fermionic holoraumy tensor 
is the appropriate mathematical structure to access the equivalent data to 
find the weights and roots of supersymmetical representations.

We can state this in another more concise way.  As the data of the eigenvalues 
(e.g. as in
\be
 {\bm T}_3 \, | t_1 , \,  t_2 > ~=~ t_1 \,  | t_1 , \,  t_2 >  ~~,~~
 {\bm T}_8 \, | t_1 , \,  t_2 > ~=~ t_2 \,  | t_1 , \,  t_2 >  ~~,
\ee
) of the maximal commuting generators opens the doorway to building a complete 
theory of representations for the su(3) Lie algebra, we assert it is the data 
of the fermionic holoraumy tensors (e.g. as
in
\be
 \eqalign{
[\, \rD_a ~,~ \rD_b \,]\, {\bm \Psi}{}_c{}^{\Hat I} ~=~  &\left[{ {\bm H}}
{}^{(CLS)}{}^{\Hat I}{}_{\Hat K}\right]{}_{a \, b \, c}{}^d \, {\bm \Psi}{}_d{}^{\Hat K} ~+~
\left[{ {\bm H}}{}^{\mu}{}^{(CLS)}{}^{\Hat I}{}_{\Hat K} \right]{}_{a \, b \, c}{}^d \, \pa_{\mu} \, {\bm 
\Psi}{}_d{}^{\Hat K} ~   \cr
&+~  
\left[{ {\bm H}}{}^{\mu \, \nu}{}^{(CLS)} {}^{\Hat I}{}_{\Hat K} \right]{}_{a \, b \, c}{}^d \, 
\pa_{\mu} \,   \pa_{\nu} \, {\bm  \Psi}{}_d{}^{\Hat K}  ~~,
}
\ee
) that opens the doorway to building a complete theory of representations for
4D, $\cal N$ = 1 spacetime supersymmetric theories...and conjecture this
can be extended for more complicated supersymmetrical theories.

The results in chapter three are very encouraging from the larger point of view 
of ``building up'' a weight space perspective on the representation space of 4D, 
$\cal N$ = 1 supermultiplets.  We thus propose that the five vectors
\be  { \eqalign{
{\vec w}_1 ~&=~ \fracm 1{\sqrt 3}\,  (1, \, 1, \, 1, \, 0)
~~~~~~\,~,~~~
{\vec w}_2 ~=~ \fracm 1{\sqrt 3}\,  (1, \, - 1, \,  - 1, \, 0)
~~~,~~~  \cr
{\vec w}_3 ~&=~ \fracm 1{\sqrt 3}\,  ( - 1, \,  1, \, - 1, \, 0)
~~~,~~~
{\vec w}_4 ~=~ \fracm 1{\sqrt 3}\,  ( - 1, \, - 1, \, 1, \, 0)
~~,~~  \cr
&~~~~~~~~~~~~~~~~~~ {\vec w}_5 ~=~  ( 0, \,  0, \, 0,  \, 1)  ~~,
} } \label{wghts}
\ee
define the weight space of these minimal 4D, $\cal N$ = 1 supermultiplets.  This
leads us to the conjecture that the ten differences 
\be  { \eqalign{
{\vec r}_{1-2} ~&=~ {\vec w}_1 ~-~ {\vec w}_2 ~=~  \fracm 2{\sqrt 3}\,  (0, \, 1, \, 1, \,  0)  ~~~~,~~~~
{\vec r}_{1-5} ~=~ {\vec w}_1 ~-~ {\vec w}_5 ~=~  \fracm 1{\sqrt 3}\,  (1, \, 1, \, 1, \,  - {\sqrt 3})  
~~~~~\,~~~,  \cr
{\vec r}_{1-3} ~&=~ {\vec w}_1 ~-~ {\vec w}_2 ~=~  \fracm 2{\sqrt 3}\,  (1, \, 0, \, 1, \,  0)  
~~~~,~~~~
{\vec r}_{2-5} ~=~ {\vec w}_2 ~-~ {\vec w}_5 ~=~  \fracm 1{\sqrt 3}\,  (1, \, -1, \, -1, \,  - {\sqrt 3})  
~~~~,  \cr
{\vec r}_{1-4} ~&=~ {\vec w}_1 ~-~ {\vec w}_4 ~=~  \fracm 2{\sqrt 3}\,  (1, \, 1, \, 0, \,  0)  
~~~~,~~~~
{\vec r}_{3-5} ~=~ {\vec w}_3 ~-~ {\vec w}_5 ~=~  \fracm 1{\sqrt 3}\,  (-1, \, 1, \, -1, \,  - {\sqrt 3})
~~~~,  \cr
{\vec r}_{2-3} ~&=~ {\vec w}_2 ~-~ {\vec w}_3 ~=~  \fracm 2{\sqrt 3}\,  (1, \, - 1, \, 0, \,  0) ~~,~~~~
{\vec r}_{4-5} ~=~ {\vec w}_4 ~-~ {\vec w}_5 ~=~  \fracm 1{\sqrt 3}\,  (-1, \, -1, \, 1, \,  - {\sqrt 3}) ~~~~,  \cr
{\vec r}_{2-4} ~&=~ {\vec w}_2 ~-~ {\vec w}_4 ~=~  \fracm 2{\sqrt 3}\,  (1, \, 0, \, -1, \,  0)  ~~,  \cr
{\vec r}_{3-4} ~&=~ {\vec w}_2 ~-~ {\vec w}_4 ~=~  \fracm 2{\sqrt 3}\,  (0, \, 1, \, -1, \,  0)  ~~,  \cr
} } \label{rts}
\ee
between the five vectors should define the root space of the 4D, $\cal N$ = 1 spacetime 
supersymmetry algebra.  We also note that
\be  \eqalign{
| \, {\vec r}_{1-2} \, | ~&=~ | \, {\vec r}_{1-3} \, | ~=~ | \, {\vec r}_{1-4} \, | ~=~ 
| \, {\vec r}_{2-3} \, | ~=~ | \, {\vec r}_{2-4} \, | ~=~ | \, {\vec r}_{3-4} \, | ~=~   2 \, {\sqrt {\fracm 23}}  ~~~, \cr
| \, {\vec r}_{1-5} \, | ~&=~ 
| \, {\vec r}_{2-5} \, | ~=~ | \, {\vec r}_{3-5} \, | ~=~ | \, {\vec r}_{4-5} \, | ~=~   {\sqrt 2}  ~~~,  \cr
{\vec r}_{A-5} \, \cdot  \, {\vec r}_{B-5}  ~&=~ \fracm 23 ~~,~~ {\rm {if ~A}} \, \in \, \{ \, 1, \, 2, \, 3, \, 4  \}, ~
{\rm {B}} \, \in \, \{ \, 1, \, 2, \, 3, \, 4  \}, ~ {\rm {and}} \, {\rm A} \, \ne \, {\rm B} ~~,  \cr
{\vec r}_{A-B} \, \cdot  \, {\vec r}_{C-5}  ~&=~ \fracm 43 \delta{}_{{\rm A} {\rm C}} ~-~
 \fracm 43 \delta{}_{{\rm B} {\rm C}} ~~~,  \cr
&~~~~~ {\rm {if ~A}} \, \in \, \{ \, 1, \, 2, \, 3, \, 4  \}, ~ {\rm {B}} 
\, \in \, \{ \, 1, \, 2, \, 3, \, 4  \}, ~ {\rm {C}} 
\, \in \, \{ \, 1, \, 2, \, 3, \, 4  \} ~, {\rm {and}} \, {\rm A} \, \ne \, {\rm B} ~~, \cr
{~~~~~} {\vec r}_{A-B} \, \cdot  \, {\vec r}_{C-D}  ~&=~ \fracm 43 \delta{}_{{\rm A} {\rm C}}
~-~ \fracm 43 \delta{}_{{\rm A} {\rm D}} ~-~ \fracm 43 \delta{}_{{\rm B} {\rm C}}
~+~  \fracm 43 \delta{}_{{\rm C} {\rm D}}  ~~,
 \cr
 & ~~~~ {\rm {if ~A}} \, \in \, \{ \, 1, \, 2, \, 3, \, 4  \}, ~ {\rm {B}} 
\, \in \, \{ \, 1, \, 2, \, 3, \, 4  \}, ~ {\rm {C}} 
\, \in \, \{ \, 1, \, 2, \, 3, \, 4  \}, ~ {\rm {D}} 
\, \in \, \{ \, 1, \, 2, \, 3, \, 4  \}  ~~~ \cr
 \, &~~~~ {\rm {and}} \, {\rm A} \, \ne \, {\rm B} ~,~  {\rm {and}} \, {\rm C} \, \ne \, {\rm D}
 ~~.
}  \ee
These equations inform us a four dimensional polytope with vertices described by the 
vectors ${\vec w}{}_1$, $\dots$ ${\vec w}{}_5$ possesses a three dimensional projection
(neglecting the ${\rm s}{}_{({\cal R})}$ direction) shown in Fig.\ 2 where the red links have 
a length of $2 \, {\sqrt {(2/3)}} $.  Similarly, another three dimensional projection (neglecting 
the ${\rm r}{}_{({\cal R})}$ direction) has the form of a pyramid consisting of a square base with
the (VS), (ATS), (AVS), and (TS) supermultiplets at the vertices of the base (taken in
the indicated order in a right-handed sense) together with the (CS) supermultiplet located at the
apex at a distance of $\sqrt 2$ from any of the other four supermultiplets.  Thus a 
``tetrahedron-pyramid'' is the ``SUSY crystal'' described by the minimal supermultiplets.

In order to find more support for the conjecture, it will be 
necessary to study larger representations of supersymmetry.  The way is open to explore 
this conjecture as the complex linear supermultiplet shows that the concept of fermionic 
holoraumy can be extended to include the three quantities $[{ {\bm H}}{}^{(CLS)}{}^{\Hat 
I}{}_{\Hat K}]{}_{a \, b \, c}{}^d$, $[{{\bm H}}{}^{\mu}{}^{(CLS)}{}^{\Hat I}{}_{\Hat K} ]{}_{a 
\, b \, c}{}^d $, and $[{ {\bm H}}{}^{\mu \, \nu}{}^{(CLS)} {}^{\Hat I}{}_{\Hat K} ]{}_{a \, b \, 
c}{}^d$.

Let us emphasize we are ignoring the role of `height' in the considerations of more
general adinkras different from valises.  This is only done for the sake of convenience.

The fact that the complex linear supermultiplet possesses twelve fermionic degrees 
of freedom and twelve bosonic degrees of freedom suggests the possibility of constructing
a $\Hat {\cal G}$ type representation space metric in the space of irreducible 12-12
supermultiplets.  There is only known to be two other such supermultiplet, namely 4D, 
$\cal N$ = 1 minimal \cite{minSG} and new minimal supergravity \cite{nwminSG}.  
Exploration of the ``$w$-vectors'' and ``$r$-vectors'' of these systems will be carried 
out in the future.

Going on beyond these examples, the next set of well studied 4D, $\cal N$ = 1 
irreducible supermultiplets consists of those with twenty fermionic degrees of 
freedom and twenty bosonic degrees of freedom.  There are three of these systems: 
\newline $~~~~~~~$ (a.) the non-minimal supergravity supermultiplet 
\cite{nonminSG1a,nonminSG1b,nonminSG2},
\newline $~~~~~~~$ (b.) the Ogievetsky-Sokatchev matter gravitino supermultiplet 
\cite{OSMGM1,OSMGM2}, and
\newline $~~~~~~~$ (c.) the de Wit-van Holten matter gravitino supermultiplet 
\cite{dWvHMGM}. \vskip.05in
The construction of these putative $\Hat {\cal G}$ type representation space metrics
will be guided by an underlying adinkra construction as was the case that led to the 
formula in (\ref{GdGET2}).  Our previous experience leads us to expect that the fermionic 
holoraumy shown for the 0-brane valise formulation (\ref{VHVH}) will almost
immediate suggest the adinkra-based one.

An even further extension consists of analysis of the two families of higher spin 4D, 
$\cal N$ = 1 supermultiplets that form towers where any arbitrary spin can be included.  
Our previous work \cite{HyrSpnFNL} has discerned that for integer values of the 
superspin $Y$ = $s$ and for each 4D, $\cal N$ = 1 supermultiplet containing 
a propagating field of spin $s$, the number of degrees of freedom is $8 s^2 + 8 s + 
4$.  This same previous work \cite{HyrSpnFNL} has discerned that for one-half 
times odd integer values of the superspin $Y$ = $s + \fracm 12$ and for each 4D, 
$\cal N$ = 1 supermultiplet containing a propagating field of spin $s + \fracm 12$, 
there are two such supermultiplets.  One has the number of degrees of freedom 
equal to $8 s^2 + 8 s + 4$ and the other has the number of degrees of freedom 
equal to $8 s^2  + 4$.

The most encouraging likely outcome is that all these supermultiplets possess 
4D, $\cal N$ = 1 fermionic holoraumy tensors that are similar in form to that given 
in (\ref{HolG1}).  The reason for this is once more simple dimensional analysis...i.e. 
the engineering dimensions of the fermions in these supermultiplets is exactly the 
same as that of the complex linear supermultiplet.

In closing, let us note that it was the study of adinkras in some of our previous
works \cite{adnkholor1,adnkholor2} which suggested that the fermionic holoraumy 
matrices calculated in the context of four dimensional $\cal N$ = 1 supermultiplets 
contained the data required to build a weight space for these supermultiplets.  As 
our observations have been similar to a phenomenological study of mathematical 
observations, we have no general theorems for their validity beyond the context we 
have examined thus far.  The main purpose of this current work is to explore the 
extension of these concepts to representations larger than those discussed in this 
chapter three and previous work.  However, it should also be noted how subtle has 
been the influence of the adinkra concept on this work.  

There is not a single four dimensional calculation in this work that depends on the 
presence of adinkras, which remain ``hidden'' (much like color in hadronic physics) 
but powerfully influence the mathematical structure of the four dimensional
results.

Thus, to the italicized question at the end of chapter two our response is that the
required data from the supersymmetry transformation laws resides in the
fermionic holoraumy tensors $[{ {\bm H}}{}^{({\widehat {\cal R}})}{}^{\Hat I}{}_{\Hat K}]{}_{a 
\, b \, c}{}^d$, $[{{\bm H}}{}^{\mu}{}^{({\widehat {\cal R}})}{}^{\Hat I}{}_{\Hat K} ]{}_{a \, b \, c}
{}^d $, and $ [{ {\bm H}}{}^{\mu \, \nu}{}^{({\widehat {\cal R}})} {}^{\Hat I}{}_{\Hat K} ]{}_{a \, 
b \, c}{}^d$ for any 4D, $\cal N$ = 1 superfield representation $({\widehat {\cal R}})$.

 \vspace{.05in}
 \begin{center}
\parbox{4in}{{\it ``Never discourage anyone...who continually makes 
progress, no matter how slow.'' \\ ${~}$ 
 \\ ${~}$ 
\\ ${~}$ }\,\,-\,\, Plato}
 \parbox{4in}{
 $~~$}  
 \end{center}
 
  \noindent
{\bf Acknowledgements}\\[.1in] \indent
This work was partially supported by the National Science Foundation grant 
PHY-1315155.  Additional acknowledgment is given by W.\ C.\ , A.\ D.\ , I.\ F.\ ,
S.\ H.\ , T.\ L-B, D.\ L.\ , K.\ M.\ , V.\ M.\ , M.\ O.\ , S.\ R.\ , D.\ S.\ , and A.\ V.\
for their participation in the 2016 SSTPRS (Student Summer Theoretical Physics 
Research Session) program at the University of Maryland Center for String \& 
Particle Theory  (CSPT).  S.J.G. acknowledges the generous support of the 
Provostial Visiting Professorship Program and the Department of Physics at 
Brown University for the very congenial and generous hospitality during the 
period of this work and his research is also supported in part by CSPT.  
Finally, we thank V.A. Korotkikh for providing on-line hosting of codes
used in this work.

\newpage
$$~~$$

\end{document}